\documentclass{aa}  
\usepackage{txfonts}
\usepackage{hyperref}

\usepackage{graphicx}	
\usepackage{amsmath}	
\usepackage{amssymb}	

\newcommand{\chandra}{\textit{Chandra}\,}
\newcommand{\angstrom}{\mbox{\normalfont\AA}}

\begin{document} 
\title{\textit{Chandra} reveals a luminous Compton-thick QSO powering a $Ly\alpha$ blob in a $z=4$ starbursting protocluster}
\titlerunning{\chandra observations of a starbursting protocluster at $z=4$}
\authorrunning{F. Vito et al.}
   \author{F. Vito\thanks{fvito.astro@gmail.com}\inst{1,2,3} \and
  	 W.N. Brandt\inst{4,5,6} \and
   	 B.D. Lehmer \inst{7} \and
   	 C. Vignali\inst{8,9}\and
           F. Zou\inst{4,5} \and
   	 F.E. Bauer\inst{2,10,11}\and
   	        M. Bremer\inst{12}\and
   	 R. Gilli\inst{9}\and
   	 R.J. Ivison\inst{13}\and
      C. Spingola\inst{8,14}
          }
 \institute{Scuola Normale Superiore, Piazza dei Cavalieri 7, 56126 Pisa, Italy
 	\and
 	Instituto de Astrof\'isica and Centro de Astroingenieria, Facultad de F\'isica, Pontificia Universidad Cat\'olica de Chile, Casilla 306, Santiago 22, Chile
         \and
Chinese Academy of Sciences South America Center for Astronomy, National Astronomical Observatories, CAS, Beijing 100012, China
\and
Department of Astronomy \& Astrophysics, 525 Davey Lab, The Pennsylvania State University, University Park, PA 16802, USA 
\and
Institute for Gravitation and the Cosmos, The Pennsylvania State University, University Park, PA 16802, USA \and
Department of Physics, The Pennsylvania State University, University Park, PA 16802, USA
\and
Department of Physics, University of Arkansas, 226 Physics Building, 825 West Dickson Street, Fayetteville, AR 72701, USA
\and
Dipartimento di Fisica e Astronomia, Università degli Studi di Bologna, via Gobetti 93/2, 40129 Bologna, Italy
\and
INAF – Osservatorio di Astrofisica e Scienza dello Spazio di Bologna, Via Gobetti 93/3, I-40129 Bologna, Italy 
\and
Millennium Institute of Astrophysics (MAS), Nuncio Monseñor Sótero Sanz 100, Providencia, Santiago, Chile
\and
Space Science Institute, 4750 Walnut Street, Suite 205, Boulder, Colorado, 80301, USA
\and
School of Physics, University of Bristol, Tyndall Avenue, Bristol BS8 1TL
\and
European Southern Observatory, Karl-Schwarzschild-Strasse 2, D-85748 Garching, Germany 
\and
INAF -- Istituto di Radioastronomia, Via Gobetti 101, I--40129, Bologna, Italy
  }

 \date{}
\abstract
 {Galaxy clusters in the local universe descend from high-redshift overdense regions known as protoclusters. The large gas reservoirs and high rate of galaxy interaction in protoclusters are expected to enhance star-formation activity and trigger luminous supermassive black-hole accretion in the nuclear regions of the host galaxies.} 
   {We investigated the active galactic nucleus (AGN) content of a gas-rich and starbursting protocluster at $z=4.002$, known as the Distant Red Core (DRC). In particular, we search for luminous and possibly obscured AGN in 13 identified members of the structure, and compare the results with protoclusters at lower redshifts. We also test whether a hidden AGN can power the $Ly\alpha$ blob (LAB) detected with VLT/MUSE in the DRC.  }
  {We observed all of the identified members of the structure with 139 ks of \chandra ACIS-S imaging. Being less affected by absorption than optical and IR bands, even in the presence of large column densities of obscuring material, X-ray observations are the best tools to detect ongoing nuclear activity in the DRC galaxies.}
   {We detect obscured X-ray emission from the two most gas-rich members of the DRC, named DRC-1 and DRC-2. Both of them are resolved into multiple interacting clumps in \mbox{high-resolution} Atacama Large Millimeter Array (ALMA) and \textit{Hubble} Space Telescope (HST) observations. In particular, DRC-2 is found to host a luminous (\mbox{$L_{2-10\,\mathrm{keV}}\approx3\times10^{45}\,\mathrm{erg\,s^{-1}}$} ) Compton-thick (\mbox{$N_H\gtrsim10^{24}\,\mathrm{cm^{-2}}$}) quasar (QSO) candidate, comparable to the most luminous QSOs known at all cosmic times. The AGN fraction among DRC members is consistent with results found for lower redshift protoclusters. However, X-ray stacking analysis reveals that supermassive black hole (SMBH) accretion is likely also taking place in other DRC galaxies that are not detected individually by \textit{Chandra}.  }
   {The luminous AGN detected in the most gas-rich galaxies in the DRC and the widespread SMBH accretion in the other members, which is suggested by stacking analysis, point toward the presence of a strong link between large gas reservoirs, galaxy interactions, and luminous and obscured nuclear activity in protocluster members. The powerful and obscured QSO detected in DRC-2 is likely powering the nearby LAB detected with VLT/MUSE, possibly through photoionization; however, we propose that the diffuse $Ly\alpha$ emission may be due to gas shocked by a massive outflow launched by DRC-2 over a $\approx10$ kpc scale.}


\maketitle
%
\section{Introduction}\label{intro}

Galaxy evolution is widely recognized to be strongly affected by environment. Galaxy growth at high redshift is enhanced in high-density regions \citep[e.g.,][]{Kauffmann96,DeLucia96}, causing the most massive and evolved spheroids in the local universe to reside in galaxy clusters \citep[e.g.,][]{Smith08}. Present-day galaxy clusters are descendants of these overdense regions, known as ``protoclusters,'' at $z \gtrsim 2$ (\citealt{Overzier16}, and references therein). 
The number of known protoclusters, and thus our knowledge of the early phases of cluster formation and evolution, is still limited. Different approaches have been used to select such structures.
Wide-field spectroscopic surveys allowed the detection of protoclusters as 3D overdensities of galaxies \citep[e.g., ][]{Steidel00,Steidel05, Cucciati14}. A similar approach adopts photometric preselection (e.g., exploiting the Lyman dropout technique or high-quality photometric redshifts) of galaxy candidates at a given redshift in multiwavelength surveys, and subsequent spectroscopic follow-up \citep[e.g.,][]{Chiang14, Toshikawa16}. At high redshift, narrow-band imaging has been used to discover overdensities of $Ly\alpha$ emitters \citep[][]{Ouchi05, Higuchi19}. Finally, galaxy overdensities have been detected around objects expected to trace highly biased regions of the universe, such as powerful radio galaxies \citep[e.g.,][]{Pentericci00,Venemans02,Overzier08, Gilli19}, QSOs up to $z\approx6.3$ (e.g., \citealt{Balmaverde17}), and $Ly\alpha$ blobs \citep[LABs; e.g.,][]{Prescott08,Badescu17}. All of these methods rely on the detection of optical and UV continuum emission and emission lines in protocluster galaxies.

Since galaxy and supermassive black hole (SMBH) growth are strongly related (\citealt{Kormendy13}, and references therein), investigating the effects of environment on SMBH growth at high redshift is necessary to obtain a complete picture (see, e.g., \citealt{Brandt15}, and references therein). In addition, nuclear activity produces both mechanical and radiative feedback, which is necessary to explain several observational findings, such as the possible enhancement of star-formation activity in protocluster members \citep[e.g.,][]{Gilli19}, the efficient quenching of star formation in clusters (especially in the most massive galaxies), and the presence of shocked hot gas, detected as ``cavities'' in the cluster extended \mbox{X-ray} emission (e.g., \citealt{Fabian12, Overzier16}, and references therein).

Several studies in the X-ray band, although limited by poor source statistics, have found enhanced active galactic nucleus (AGN) activity in massive protoclusters at $2 \lesssim z \lesssim3$ with respect to the field environment \citep[e.g.,][]{Pentericci02,Lehmer09b,Lehmer13,Digby-North10}. Strikingly, this is the opposite of what is found for galaxy clusters in the local universe \citep[e.g.,][]{Kauffmann04,Eastman07,Martini06,Martini09,Martini13,Ehlert14}. This behavior is similar to the reversal of the relation between star-formation rate density and environment from the distant to the local universe \citep[e.g.,][]{Elbaz07, Brodwin13}. A possible explanation is that unvirialized, overdense regions at high redshift provide both the large reservoirs of cold gas and the mechanisms (mergers and galaxy encounters) to drive this gas into the SMBH potential well and trigger nuclear accretion, while SMBH accretion is prevented in the virialized structures at low redshift \citep[e.g.,][]{vanBreukelen09}. Pushing the study of AGN in protoclusters to earlier cosmic times would thus provide a key insight into the connection between SMBHs, host galaxies, and their environment.

At high redshift, the availability of large gas reservoirs is also expected to trigger intense \mbox{star-formation} activity in the galaxies residing in overdense regions, as is often observed in protocluster candidates \citep[e.g.,][]{Capak11,Umehata15,Umehata19,Kubo19}. Starbursting galaxies, often identified as dusty star-forming galaxies (DSFGs; e.g., \citealt{Hodge20}) or sub-mm galaxies (SMGs), represent a short but key phase in galaxy evolution, possibly resulting from ``wet" (i.e., gas-rich) merger events.  
In the framework of  BH-galaxy coevolution scenarios \citep[e.g.,][]{Hopkins08}, luminous AGN are also thought to be triggered under the same conditions, and eventually to shut down both nuclear accretion and star-formation activity in the host galaxies through negative feedback. While some luminous and obscured AGN have been detected in SMGs \citep[e.g.,][]{Alexander16,Ivison19} or other starbursting galaxies \citep[e.g.,][]{Tsai15}, the prevalence of nuclear accretion in SMGs is still a debated issue \citep[e.g.,][]{Alexander05,Wang13_1}.

Usually, DSFGs are very faint in the optical/UV bands, due to strong dust extinction, and are generally  missed by widely used protocluster selection methods. However, recently,
 bright FIR-to-mm sources have been resolved with ALMA into multiple DSFGs at high redshift ($z\gtrsim4$), pinpointing the positions of massive, star-forming and gas-rich galaxy overdensities \citep[e.g.,][]{Oteo18,Miller18,Hill20}. This approach allowed the discovery of  one of the most extreme protoclusters in terms of total mass, star-formation rate, and gas mass known at high redshift (\mbox{$z=4.002$}, i.e., the redshift of the brightest member), referred to as the Distant Red Core (DRC; RA = 00:42:23.8, Dec = $-$33:43:34.8), as discussed by \cite{Ivison16}, \cite{Fudamoto17}, \cite{Lewis18}, \cite{Oteo18}, \cite{Long20}, and \cite{Ivison20}.
The DRC was first selected by \cite{Ivison16} as a high-redshift DSFG candidate on the basis of its very red \textit{Herschel} colors in the H-ATLAS survey \citep{Eales10}. Subsequent LABOCA and ALMA observations resolved this source into several individual galaxies, identified as DRC-1 to DRC-11, in descending order of 2 mm continuum brightness, in the inner core of the structure, over an area of $\approx260\,\mathrm{kpc} \times 310\,\mathrm{kpc}$, and sources C to G in the external regions. 

Currently, 13 of these sources have been spectroscopically confirmed with ALMA to be members of the same structure at $z=4$ (\citealt{Oteo18,Ivison20}) via detection of up to five emission lines ($^{12}$CO(2-1), $^{12}$CO(4-3), $^{12}$CO(6-5), [C I](1-0), H$_2$(2$_{11}-2_{02}$)). The identified members are all of the core galaxies except for DRC-5, plus three galaxies in the external regions of the protocluster (i.e., Sources C, D, and E). Several continuum and line emitters have been detected in the fields of sources C-G but remain unidentified, and possibly some of them belong to the same $z=4$ overdensity. In particular, a second CO(4--3) line emitter was detected at $\approx3$ arcsec and $\Delta v\approx1700\,\mathrm{km\,s^{-1}}$ from source D. It is currently unclear if this is another member of the DRC aligned with source D, or if it traces an outflow launched by that source. For the purpose of this paper, we will consider the two line emitters in source D as a single structure member. 
The core of the structure is characterized by a DSFG surface density that is $10\times$ higher than SSA22 \citep{Umehata15} at $z=3.09$, which is one the most massive structures known at high redshift \citep{Steidel00}). Radio observations ($5.5-28$ GHz) identified DRC-6 as a radio galaxy, with a radio flux 50$\times$ brighter than that expected from star formation \citep{Oteo18}. A high-resolution ($\approx0.12$ arcsec) 870$\mu m$ ALMA image of the brightest DSFG (DRC-1) reveals it is composed of three distinct and likely interacting star-forming clumps \citep{Oteo18}. Similarly, recent HST/WFC3 observations in the F125W band resolved several DRC members into multiple star-forming and likely interacting clumps on scales of a few kpc \citep{Long20}.

Individual star-formation rates of confirmed protocluster members, derived from the FIR or CO(4--3) line luminosities by \cite{Oteo18} and \cite{Ivison20}, and from spectral energy distribution (SED) fitting by \cite{Long20}, are in the range $SFR=50-3000\,\mathrm{M_\odot\,yr^{-1}}$. The total $SFR\approx6000-7400\,\mathrm{M_\odot\,yr^{-1}}$ is the highest among known protoclusters at $z=2.5-5.3$ \citep{Oteo18,Hill20}, with $\approx70\%$ attributed to the three brightest member galaxies (i.e., DRC-1, DRC-2, and DRC-3). It is likely that the DRC has been caught during a phase of fast and efficient stellar-mass building, not yet quenched by AGN feedback. This is also supported by the large estimated molecular gas mass of the structure ($M_{H_2}>10^{12}\,M_\odot$; \citealt{Long20}). The estimated total mass of the protocluster  $M_{tot}\approx10^{14}\,M_\odot$ makes the DRC a likely progenitor of a Coma-like cluster in the local universe \citep{Long20}. The vigorous collective SFR of the DRC is most likely fueled by a continuous inflow of gas \citep[e.g.,][]{Umehata19}. The same gas reservoir is also expected to trigger efficient, fast, and obscured SMBH growth typical of the peak phases of SMBH growth as predicted by several theoretical models \citep[e.g.,][]{Menci03,Hopkins06a}.

A VLT/MUSE observation of the DRC reveals the presence of an LAB (with linear size $\approx60$ kpc) close to DRC-2 \citep{Oteo18}. Similar objects are often found in protoclusters \citep[e.g.,][]{Overzier16}, and are signposts of large amounts of neutral gas. 
The powering sources of LABs have been often identified with the photoionizing emission from a close ionizing source (e.g., a QSO) \citep[e.g.,][]{Geach09,Overzier13,Hennawi15}, resonant scattering of $Ly\alpha$ photons emitted by a central source \citep[e.g.,][]{Borisova16}, shocks \citep[e.g.,][]{Taniguchi00, Mori04}, or ``cooling radiation'' following collisional ionization during the gravitational collapse of the gas \citep[e.g.,][]{Haiman00}. 
None of the protocluster members were detected as $Ly\alpha$ emitters in the MUSE data \citep{Oteo18}.

In this paper, we present the results of a 139 ks \chandra observation of the DRC, aimed at investigating its AGN content. We compare it with similar structures at lower redshift and assess whether the source of energy powering the LAB is a buried AGN. 
Uncertainties and limits are reported at a 68\% confidence level, unless otherwise noted. We adopt a flat $\mathrm{\Lambda}$CDM cosmology with $H_0=67.7\,\mathrm{km\,s^{-1}}$ and $\Omega_m=0.307$ \citep{Planck16}. At $z=4$, $1$ arcsec corresponds to $\approx7.1$ kpc.

\section{\chandra data reduction and analysis}
\begin{figure}
	\begin{center}
		\hbox{
			\includegraphics[width=90mm,keepaspectratio]{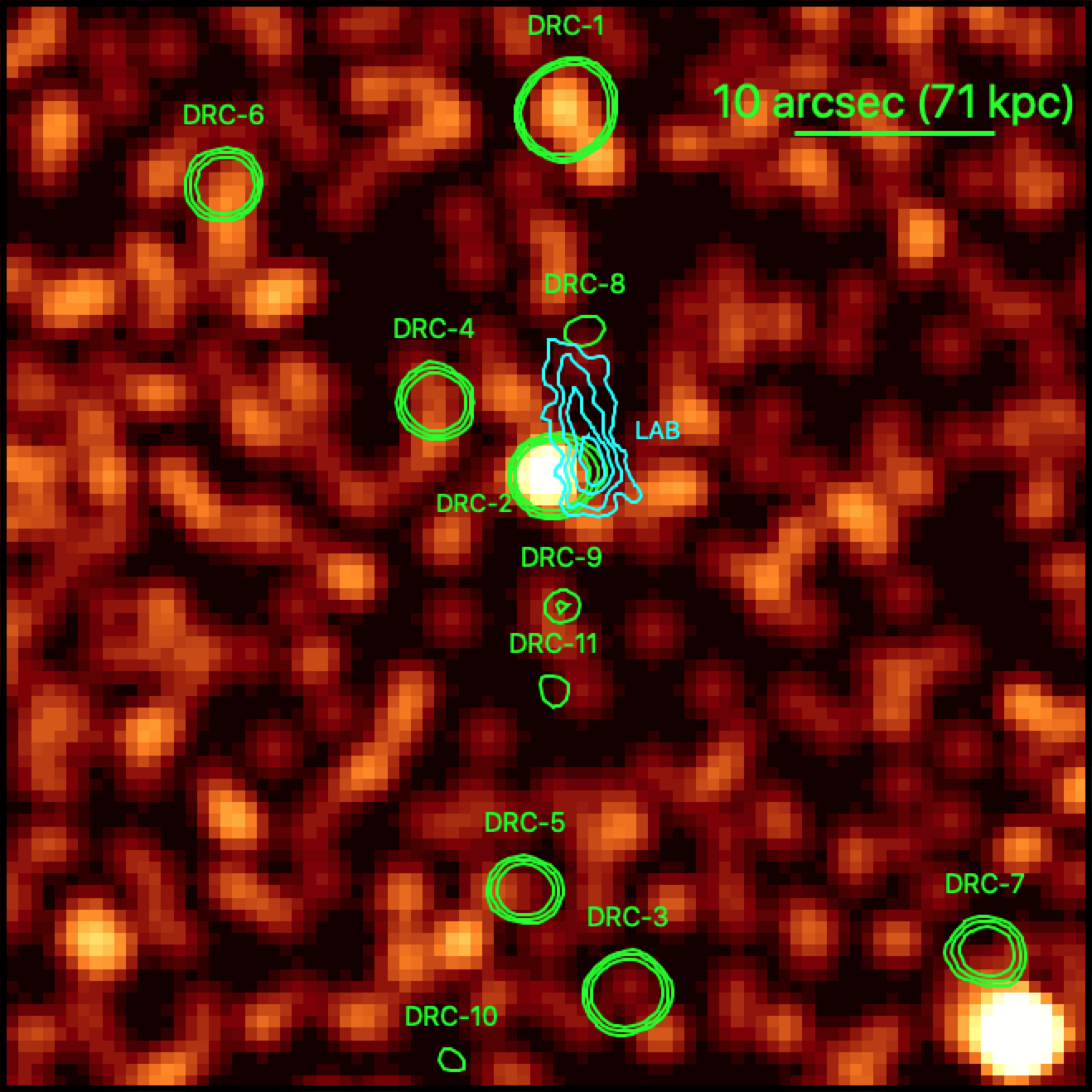} 
		}
	\end{center}
	\caption{Smoothed $45''\times45''$ \chandra full-band image of the DRC core. The green contours ($3\sigma$, $5\sigma$, $9\sigma$) mark the 2 mm continuum emission of the ALMA sources with $\approx2"$ angular resolution, labeled as in \citet{Oteo18}. The cyan contours ($5\sigma$, $7\sigma$, $9\sigma$, $11\sigma$) show the continuum-subtracted LAB emission in a velocity range $[-1300,800]\, \mathrm{km\,s^{-1}}$, centered on the $Ly\alpha$ wavelength at $z=4.002$.  This velocity range encompasses most of the $Ly\alpha$ line profile of the LAB \citep[see][]{Oteo18}. The entire image is covered by MUSE observations.
	}\label{Fig_field}
\end{figure}

We observed the DRC with \chandra\, for a total of 139 ks, split into 3 observations (see Tab.~\ref{Tab_X-ray_obs} for details).
We used the back-illuminated ACIS S3 CCD to cover all of the spectroscopically identified members of the structure. The core of the DRC has been placed at the aimpoint of the observations, where the \chandra sensitivity is highest.
We reprocessed the \chandra\, observations with the \textit{chandra\_repro} script in CIAO 4.11 \citep{Fruscione06},\footnote{\url{http://cxc.harvard.edu/ciao/}} using CALDB v4.8.4,\footnote{\url{http://cxc.harvard.edu/caldb/}} setting the option \textit{check\_vf\_pha=yes}, since observations were taken in very faint mode. In order to correct the astrometry of each observation, we performed source detection with the \textit{wavdetect} script with a significance threshold of $10^{-6}$, and matched the positions of the detected point sources with the USNO-B1 optical source catalog (which has a typical $0.2''$ astrometric accuracy; \citealt{Monet03}) using the \textit{wcs\_match} and \textit{wcs\_update} tools. Finally, we merged the individual observations with the \textit{reproject\_obs} tool, and derived merged images and exposure maps. Fig.~\ref{Fig_field} presents the \chandra 0.5--7 keV image of the DRC. Spectra, response matrices, and ancillary files were extracted from individual pointings using the \textit{specextract} tool and were added using the \textit{mathpha}, \textit{addrmf}, and \textit{addarf} HEASOFT tools\footnote{\url{https://heasarc.gsfc.nasa.gov/docs/software/heasoft/}}, respectively, weighting by the individual exposure times. Ancillary files, which are used to derive fluxes and luminosities, were aperture corrected.

We assessed the detection significance for each member of the structure in three energy bands: 0.5--2 keV, 2--7 keV, and 0.5--7 keV, which we refer to as the soft, hard, and full bands, respectively. We stress that at $z=4$ these bands sample rest-frame energies up to $35$ keV, allowing us to detect hard X-rays, which are not absorbed even by large (i.e., up to highly Compton-thick) column densities of obscuring material.
We computed the detection significance with the binomial no-source probability \mbox{\citep{Weisskopf07,Broos07}}
\begin{equation}
P_B(X\geq S)= \sum_{X=S}^{N}\frac{N!}{X!(N-X)!}p^X(1-p)^{N-X},
\end{equation}
where $S$ is the total number of counts in the source region in the considered energy band, $B$ is the total number of counts in the background region, \mbox{$N=S+B$}, and $p=1/(1+BACKSCAL)$, with $BACKSCAL$ being the ratio of the background and source region areas. 
We consider a source to be detected if $(1-P_B)>0.99$. This threshold is often used for assessing the detection significance of faint sources at predetermined positions \citep[e.g.,][]{Vito19b}.
We used circular extraction regions with $R=1''$ for evaluating the source counts S, and nearby regions free of contaminating sources for background evaluation.

\begin{table}
	\caption{Summary of the \chandra\, observations of the DRC.}
	\begin{tabular}{ccccccccc} 
		\hline
		\multicolumn{1}{c}{{ OBSID}} &
		\multicolumn{1}{c}{{ Start date }} &
		\multicolumn{1}{c}{{ $T_{exp}$ [ks]}} \\ 
21463 &2019-09-13 &57\\
22843 &2019-09-15 &59\\
22844 &2019-09-16 &23\\
		\hline
	\end{tabular} \label{Tab_X-ray_obs}\\
\end{table}

\section{Results}\label{results}

Among the 13 spectroscopically confirmed protocluster members, DRC-1 and DRC-2 are detected in our \chandra observations (\S~\ref{DRC-1} and \ref{DRC-2}). We performed stacking analysis on the remaining 11 undetected galaxies to study their average X-ray emission (\S~\ref{stack} and Appendix~\ref{Appendix_undetect}).
We also detected in the X-rays three ALMA sources in the DRC field that are not yet identified spectroscopically (Appendix~\ref{Appendix_unident}).

\subsection{X-ray emission from DRC-1}\label{DRC-1}

\begin{figure}
	\begin{center}
		\hbox{
			\includegraphics[width=90mm,keepaspectratio]{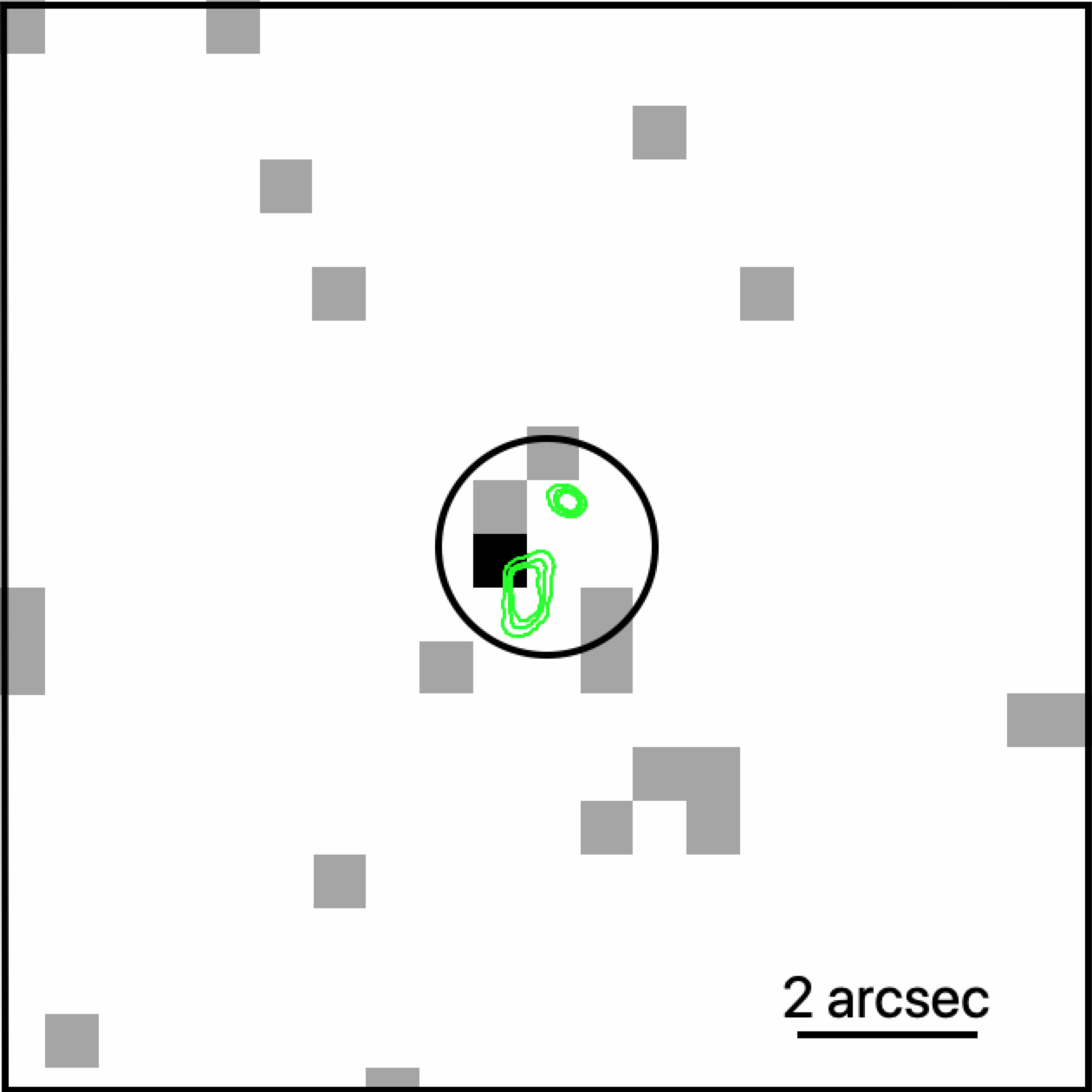} 
		}
	\end{center}
	\caption{\chandra full-band image of DRC-1. The black circle is the region used for X-ray photometry evaluation. Green contours ($3\sigma$, $7\sigma$, $11\sigma$) mark the three ALMA sources identified with a high-resolution ($0.12''$) 870 $\mu$m observation. Even high-resolution ALMA imaging can barely resolve the two southern sources (see \citealt{Oteo18}). }\label{Fig_DRC1}
\end{figure}

DRC-1 is significantly detected ($1-P_B>0.999$) in the full band (Fig.~\ref{Fig_DRC1}) and hard band with $4.4^{+ 2.6}_{-1.9}$ and $3.6^{+ 2.4}_{-1.7}$ net counts, respectively, while it is not detected in the soft band  (\mbox{$<2.2$} net counts). The hardness ratio\footnote{The hardness ratio is defined as $HR=\frac{(H-S)}{(H+S)}$, where $S$ and $H$ are the net counts in the soft band and hard band, respectively. It can be used to derive a simple description of the spectral shape of an X-ray source. } $HR>0.29$ corresponds to an effective photon index $\Gamma_{eff}<0.55$. 
If we assume an intrinsic power-law spectrum with $\Gamma=1.9$, the $HR$ limit implies a lower limit on the obscuring column density of \mbox{$N_H\geq7\times10^{23}\,\mathrm{cm^{-2}}$}. This spectral model implies fluxes \mbox{$F_{0.5-2\,\mathrm{keV}}<1.4\times10^{-16}\,\mathrm{erg\,cm^{-2}\,s^{-1}}$} and \mbox{$F_{2-7\,\mathrm{keV}}=7.6^{+5.0}_{-3.5}\times10^{-16}\,\mathrm{erg\,cm^{-2}\,s^{-1}}$}
in the soft and hard bands, respectively. Correcting for a column density of \mbox{$N_H\geq7\times10^{23}\,\mathrm{cm^{-2}}$}, we estimate a rest-frame X-ray luminosity of \mbox{$L_{2-10\,\mathrm{keV}}\geq1.2^{+0.7}_{-0.5}\times10^{44}\,\mathrm{erg\,s^{-1}}$}.

Despite the very limited number of detected photons, we extracted the X-ray spectrum and fitted it with a simple absorbed power-law model (\textit{phabs*zphabs*pow} in XSPEC,\footnote{XSPEC v.12.9 \citep{Arnaud96}, \url{https://heasarc.gsfc.nasa.gov/xanadu/xspec/}. We fitted the X-ray spectrum using the modified Cash statistics \citep[][\url{https://heasarc.gsfc.nasa.gov/xanadu/xspec/manual/XSappendixStatistics.html}]{Cash79}} where \textit{phabs} accounts for Galactic absorption; \citealt{Kalberla05}), fixing $\Gamma=1.9$. We used the W-statistic,\footnote{\url{https://heasarc.gsfc.nasa.gov/xanadu/xspec/manual/XSappendixStatistics.html}} which extends the \cite{Cash79} statistic in the case of background-subtracted data. The best-fitting model returns a column density \mbox{$N_H=1.8^{+3.2}_{-0.9}\times10^{24}\,\mathrm{cm^{-2}}$}, fluxes of \mbox{$F_{0.5-2\,\mathrm{keV}}=3^{+2}_{-2}\times10^{-17}\,\mathrm{erg\,cm^{-2}\,s^{-1}}$} and \mbox{$F_{2-7\,\mathrm{keV}}=1.0^{+0.9}_{-0.5}\times10^{-15}\,\mathrm{erg\,cm^{-2}\,s^{-1}}$}, and an absorption-corrected luminosity \mbox{$L_{2-10\,\mathrm{keV}}=2.9^{+2.3}_{-1.6}\times10^{44}\,\mathrm{erg\,s^{-1}}$}. According to these values, which are in line with those estimated from the hardness ratio, DRC-1 hosts a heavily obscured, and possibly Compton-thick, AGN with moderate luminosity.

DRC-1 is the confirmed protocluster member with the highest molecular gas mass ($M_{H_2}\approx9\times10^{11}\,M_\odot$), and one of the most massive in terms of stellar mass \citep[$M_{*}\approx2\times10^{11}\,M_\odot$;][]{Long20}. 
Moreover, it is resolved into three star-forming clumps by a high-resolution ($0.12''$) ALMA observation at 870 $\mu$m \citep[][see green contours in Fig.~\ref{Fig_DRC1}]{Oteo18} and rest-frame UV HST observations \citep{Long20}. Both the availability of a large gas reservoir and possible galaxy interaction are thought to trigger luminous AGN activity \citep[e.g.,][]{Vito14b,Donley18,Weigel18}. SMGs, in particular, are both gas rich and often associated with galaxy mergers \citep[e.g.,][and references therein]{Alexander12, Narayanan15}, as in the case of DRC-1. Therefore, it is not surprising to find luminous and obscured AGN among them \citep[e.g.,][]{Ivison98,Ivison19, Wang13_1}. Due to \chandra's angular resolution and the limited photon statistics, we are unable to identify which of the three clumps constituting DRC-1 is the X-ray source, or if more than one of them provides significant contributions to the total X-ray emission. 

DRC-1 is forming stars at a powerful rate (\mbox{$\mathrm{SFR}\approx1700-2900\,\mathrm{M_\odot\,yr^{-1}}$}; \citealt{Oteo18, Long20}). X-ray emission from high-mass X-ray binaries (HMXBs), which is proportional to SFR \citep[e.g.,][]{Bauer02, Ranalli03,Ranalli05, Basu-Zych13_2,Symeonidis14,Lehmer19}, can thus contribute significantly to the X-ray emission
detected from this source. The formation timescale of HMXBs after a star-formation event is on the order of $1-10$ Myr \citep[e.g.,][]{Fragos13,Garofali18}. To estimate the contribution of HMXBs to the X-ray emission of DRC-1, we used the \cite{Fragos13} model 269 of the evolution of $L_X$/$SFR$, which has been found by \cite{Lehmer16} to be the one that best agrees with observational
data up to $ z \approx2.5$. According to such a model, a value of \mbox{$\mathrm{SFR}=2900\,\mathrm{M_\odot\,yr^{-1}}$} at $z=4$ can produce an X-ray luminosity of \mbox{$L_{2-10\,\mathrm{keV}}=2\times10^{43}\,\mathrm{erg\,s^{-1}}$}, a factor of $>6$ times lower than the estimated luminosity of DRC-1. Moreover, XRB populations usually produce X-ray spectra softer than the emission detected from DRC-1 \citep[e.g.,][]{Mineo12,Lehmer15}.

Several caveats should be noted when using this argument. For instance, the \cite{Fragos13} theoretical models have been observationally validated only up to $z=2.5$, although they are consistent with results at $z=3-5$ \citep{Vito16}. Moreover, the physical assumptions made by \cite{Fragos13} might not be valid for sub-mm bright DSFGs. For instance, at a fixed SFR, the X-ray emission produced by XRBs is expected to decrease with increasing metallicity \citep[e.g.,][]{Basu-Zych16,Brorby16,Fornasini19, Fornasini20}. The metallicity of high-redshift DSFGs is probably higher than that of the typical galaxy population \citep[e.g.,][]{Swinbank04, Rigopoulou18, DeBreuck19}. Therefore, the XRB contribution to the X-ray luminosity has been observed to be lower in DSFGs than in typical star-forming galaxies. 
Hereafter we will consider the X-ray emission in DRC-1 to be primarily produced by an AGN.

\subsection{X-ray emission from DRC-2}\label{DRC-2}
\begin{figure}
	\begin{center}
		
		\includegraphics[width=90mm,keepaspectratio]{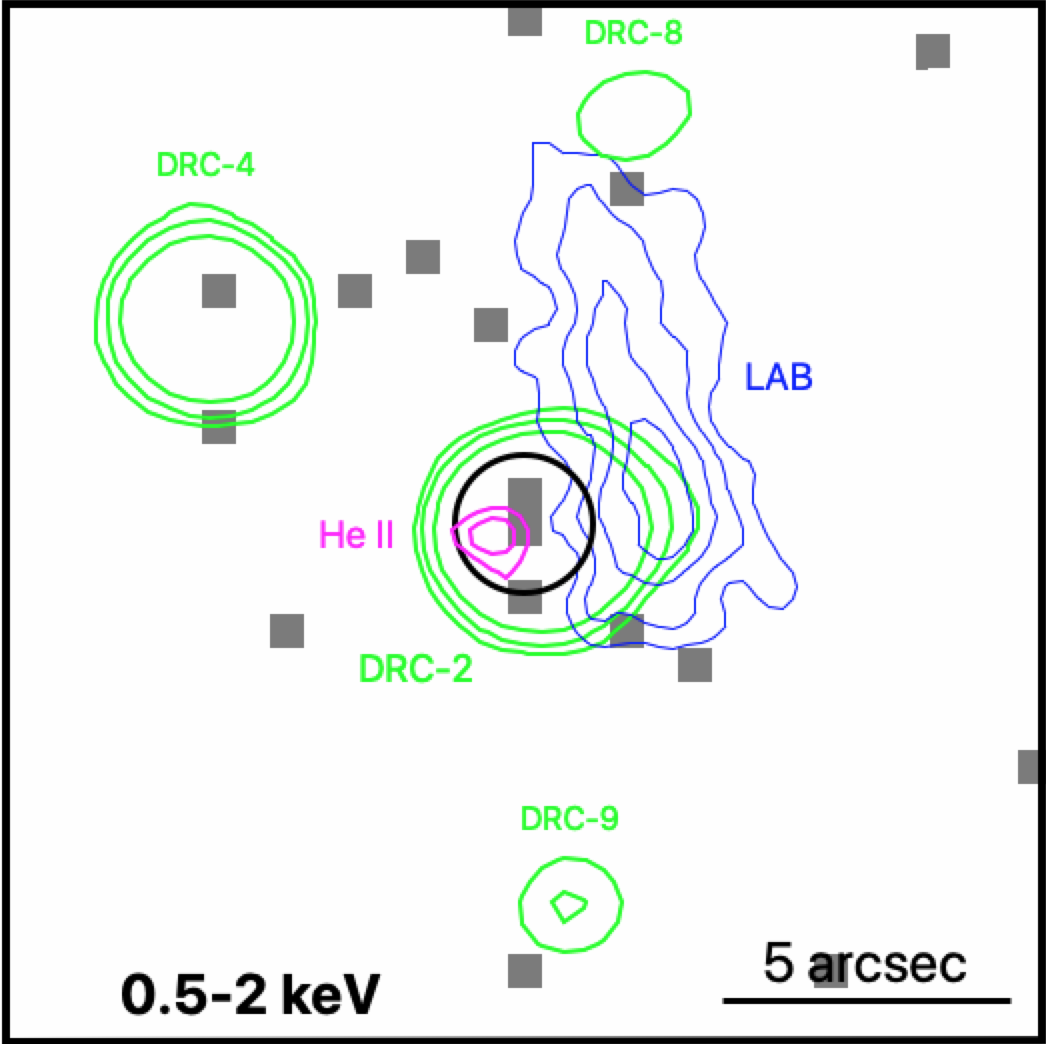}
		\includegraphics[width=90mm,keepaspectratio]{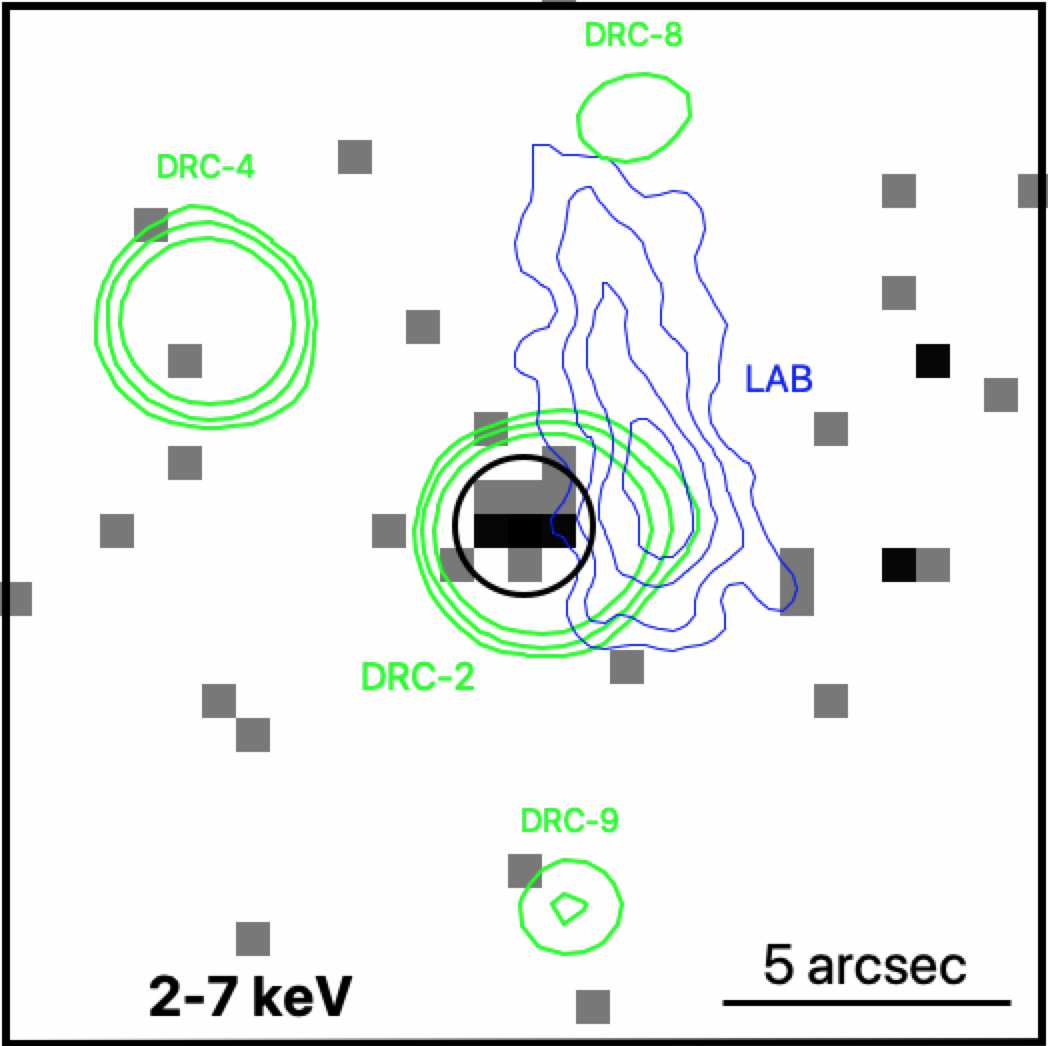} 	
	\end{center}
	\caption{\chandra soft-band (top panel) and hard-band (bottom panel) images of DRC-2. The black circle is the region used for X-ray photometry evaluation, centered on the X-ray emission. Green contours indicate the ALMA emission at 2 mm.  The continuum-subtracted emission of the $Ly\alpha$ blob (LAB) is shown with blue contours (see \S~\ref{DRC2_LAB}).  In the upper panel we also show the contours of the He II emission ($4\sigma$ and $5\sigma$) at the position of DRC-2 (see \S~\ref{DRC2_spectrum}).}   \label{Fig_DRC2}
\end{figure}

DRC-2 (Fig.~\ref{Fig_DRC2}) is detected in the full band ($14.4^{+4.2}_{-3.6}$ net counts) and hard band ($12.6^{+ 4.0}_{-3.3}$ net counts), while it is not detected in the soft band ($<3.3$ net counts).
Such relatively bright emission at high energies (10--35 keV in the rest frame) not associated with a soft-band detection is indicative of heavy obscuration. Fitting the X-ray spectrum with a simple power-law, we obtained a very hard effective photon index (i.e., $\Gamma_{eff}=-0.9^{+0.6}_{-0.7}$).

We first modeled the DRC-2 X-ray spectrum with an absorbed power-law emission (\textit{wabs*zphabs*pow} in XSPEC). Due to the limited photon counting statistics, we fixed $\Gamma{=}1.9$. The best-fitting model returns a hard-band flux \mbox{$F_{2-7\mathrm{keV}}=3.8^{+2.5}_{-2.4}\times10^{-15}\,\mathrm{erg\,cm^{-2}\,s^{-1}}$}. The intrinsic absorbing column density along the line of sight is \mbox{$N_H=6.3^{+3.3}_{-4.2}\times10^{24}\,\mathrm{cm^{-2}}$}, and the absorption-corrected rest-frame luminosity is $L_{2-10\,\mathrm{keV}}=2.0^{+1.2}_{-1.3}\times10^{45}\,\mathrm{erg\,s^{-1}}$. However, this simple spectral model may not be a good representation of the X-ray emission in the case of heavy obscuration, as additional components, such as scattered emission, become increasingly important.

Therefore, we also fit the DRC-2 X-ray spectrum with the  MYTorus model \citep{Murphy09}, which treats self-consistently both the transmission and the reflected components. Due to the limited photon counting statistics, we fixed $\Gamma{=}1.9$, the normalization of the scattered and line components to that of the transmitted component, and the inclination angle to 90 deg. 
The best-fitting model returns  \mbox{$F_{2-7\mathrm{keV}}=3.4^{+11.2}_{-2.0}\times10^{-15}\,\mathrm{erg\,cm^{-2}\,s^{-1}}$}, while we derived only a lower limit on the column density \mbox{($N_H>1.2\times10^{24}\,\mathrm{cm^{-2}}$)}.\footnote{We restricted the hard upper limit of $N_H$ in the MYTorus model from $10^{25}\,\mathrm{cm^{-2}}$ to $5\times10^{24}\,\mathrm{cm^{-2}}$, as for $N_H\gtrsim5\times10^{24}\,\mathrm{cm^{-2}}$ the column density and the power-law component normalization become completely degenerate and insensitive to the data, due to the limited photon statistics, leading to the appearance of a second minimum of the fit statistics at $N_H=10^{25}\,\mathrm{cm^{-2}}$.} The absorption-corrected rest-frame X-ray luminosity is \mbox{$L_{2-10\,\mathrm{keV}}=2.7^{+8.9}_{-1.6}\times10^{45}\,\mathrm{erg\,s^{-1}}$}. 
A pure reflection model, obtained by setting the normalization of the transmitted component to zero, returns an equally good quality fit, with intrinsic \mbox{$L_{2-10\,\mathrm{keV}}=5.3^{+7.7}_{-2.0}\times10^{45}\,\mathrm{erg\,s^{-1}}$}.

\begin{figure}
	\begin{center}
		\hbox{
			\includegraphics[width=90mm,keepaspectratio]{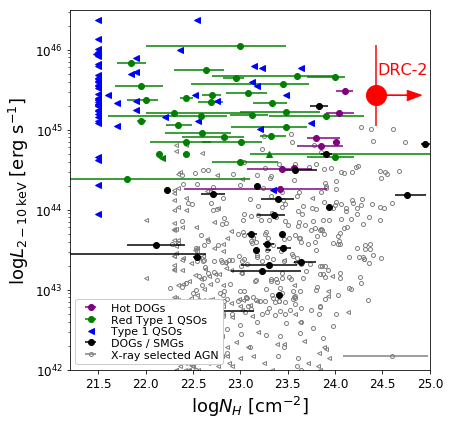} 
		}
	\end{center}
	\caption{Intrinsic X-ray luminosity versus obscuring column density for DRC-2 (red symbol) compared with several sub-populations of AGN. Only objects detected in the X-rays are included. Hot dust-obscured galaxies (Hot DOGs; purple symbols) are collected from \citet{Stern14}, \citet{Assef16}, \citet{Ricci17},\citet{Vito18b}, and \citet{Zappacosta18}. Red Type 1 QSOs (or Extremely Red QSOs, ERQs, green symbols) are from \citet{Urrutia05}, \citet{Banerji14}, \citet{Mountrichas17}, \citet{Goulding18}, and \citet{Lansbury20}. Type 1 QSOs (blue symbols) are collected from \citet[arbitrarily placed at log$N_H=21.5$]{Just07} and \citet{Martocchia17}. Dust obscured galaxies (DOGs) and SMGs (black symbols) are collected from \cite{Wang13_1}, \cite{Corral16}, and Zou et al. (submit.).  Grey symbols represent X-ray selected AGN population in the \textit{Chandra} deep field-south \citep{Li19}, with typical error bars showed in the lower right corner of the figure. Leftward pointing arrows represent upper limits on the column density.
 	}\label{Fig_DRC2_Lx_NH}
\end{figure}

In Fig.~\ref{Fig_DRC2_Lx_NH} we compare the best-fitting parameters for \mbox{DRC-2}, derived with the MYTorus model, with other sub-populations of AGN. The X-ray luminosity of DRC-2 is typical of  luminous optically selected Type 1 QSOs. However, DRC-2 is obscured by a large column density, similar to some luminous hot dust-obscured galaxies (Hot DOGs) and X-ray selected AGN with moderate luminosity.
DRC-2 is therefore one of the most luminous and distant QSOs deeply buried in Compton-thick material \citep[e.g.,][]{Vito18b}, and the first one to be found in a protocluster environment \citep[e.g.,][]{Lehmer09a,Digby-North10}. 

 DRC-2 is the second most gas-rich ($M_{H_2}\approx3\times10^{11}\,\,\mathrm{M_\odot}$) and star-forming \citep[$\mathrm{SFR}\approx1000\,\mathrm{M_\odot\,yr^{-1}}$;][]{Oteo18,Long20} member of the protocluster after DRC-1. Moreover, like DRC-1, it is resolved into multiple rest-frame UV components, suggesting ongoing galaxy interaction. These properties confirm the presence of a strong link between massive gas reservoirs, galaxy interactions, and powerful and obscured nuclear activity. Given the SFR of DRC-2, the \cite{Fragos13} model predicts that HMXBs would only produce a factor $>300$ weaker X-ray emission than the measured luminosity. The same caveats regarding XRBs discussed at the end of \S~\ref{DRC-1} hold also for DRC-2.

\subsection{Stacking results for undetected sources}\label{stack}
\begin{figure}
	\begin{center}
		
		\includegraphics[width=90mm,keepaspectratio]{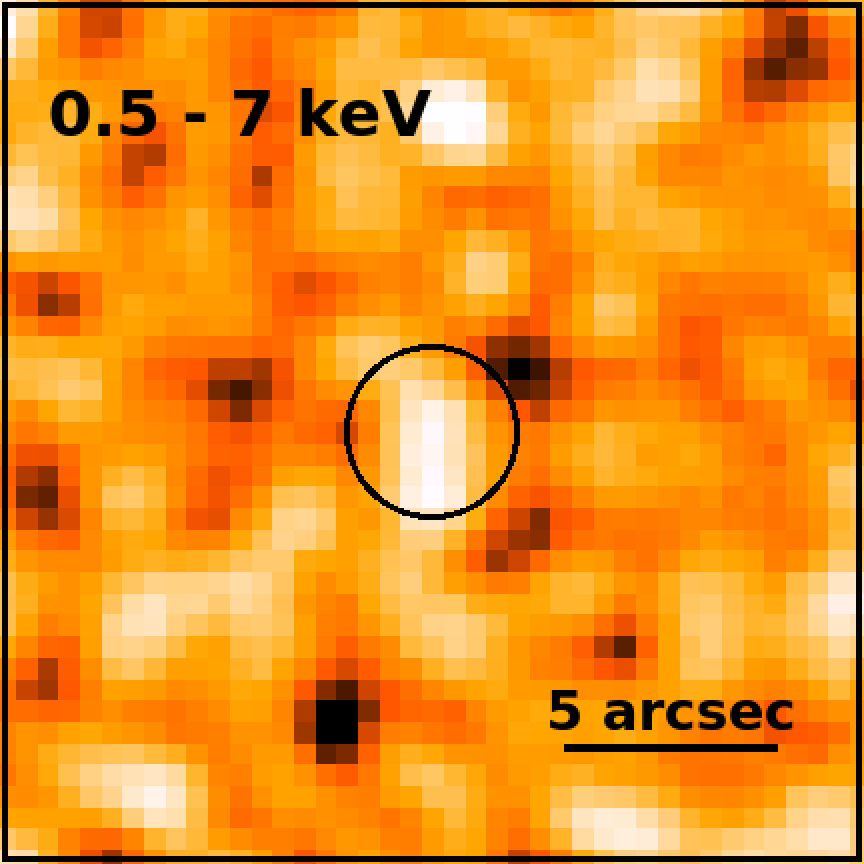}
		
	\end{center}
	\caption{Stacked and smoothed $25''\times25''$ image in the full band of the 11 individually undetected, spectroscopically confirmed members of the protocluster. }\label{Fig_stacking}
\end{figure}

Stacking analysis can be used to place constraints on the average X-ray emission of the spectroscopically confirmed members of the structure that are not detected individually by \textit{Chandra}: i.e., DRC-3, DRC-4, DRC-6, DRC-7, DRC-8, DRC-9, DRC-10, DRC-11, Source C, Source D, and Source E (see Appendix~\ref{Appendix_undetect} for individual X-ray images). First, we used the \textit{dmfilth} tool to mask field X-ray sources detected with \textit{wavdetect} at a significance of $10^{-6}$ in the full band. Then we stacked the \chandra\, emission centered on the ALMA positions with the \textit{ dmimgcalc} tool in the three standard X-ray bands. Fig.~\ref{Fig_stacking} presents the stacked image in the full band, which has an effective exposure time of $\approx1.5$ Ms.

We used a $R=1''$ circular aperture to extract the source photometry of the stacked images  (16 total counts in the full band), and annular regions with inner and outer radii of $3$ arcsec and $12$ arcsec, respectively,  to extract the background photometry. We  detected ($1-P_B=0.997$) stacked emission in the full band with $8.8^{+4.3}_{-3.7}$ net counts, while the stacked emission in both the soft and hard bands is slightly below the detection threshold. Assuming $\Gamma=1.4$ power-law emission, characteristic of the cosmic X-ray background, the stacked photometry corresponds to an average flux \mbox{$F_{0.5-7\,\mathrm{keV}}=1.2^{+0.6}_{-0.5}\times10^{-16}\,\mathrm{erg\,cm^{-2}\,s^{-1}}$} and an average rest-frame luminosity \mbox{$L_{2-10}\,\mathrm{keV}=3.6^{+1.8}_{-1.5}\times10^{43}\,\mathrm{erg\,s^{-1}}$}.
Therefore, our stacking analysis suggests that nuclear activity with low or moderate accretion rate and possibly obscured may be ongoing even among the DRC members that are not individually detected in our \chandra\, observations. We note that the reported luminosity is $\gtrsim1$ order of magnitude higher than the expected X-ray emission from XRBs \citep{Fragos13}, given the typical $SFR$ of a few $100\,\mathrm{M_\odot\,yr^{-1}}$ of the stacked galaxies \citep[e.g.,][]{Oteo18,Long20} .

While all galaxies contribute at most $\lesssim20\%$ to the stacked X-ray flux, one of them, DRC-6, is a known radio galaxy (see \S~\ref{intro} and \ref{AGN_content}) that is expected to emit in the \mbox{X-ray} band, and in fact provides one of the major contributions (3 out of the 16 total counts in the source region). We repeated the stacking analysis excluding DRC-6, and found that the significance of the stacked emission decreases to $(1-P_B)=0.984$ (i.e., slightly below the detection threshold we set for the detection of individual sources in \S~\ref{results}). This result implies that X-ray sources in the stacked galaxies, if present, are low luminosity and possibly obscured AGN, although a fractional contribution from XRBs cannot be excluded.

\section{Discussion}
The presence of luminous and obscured QSO candidates revealed by \chandra in DRC-1 and DRC-2 (see \S~\ref{DRC-1} and \S~\ref{DRC-2}) motivated us to study in more detail the physical properties of these galaxies. In \S~\ref{DRC1_SED} and \S~\ref{DRC2_SED} we fit the SED of DRC-1 and DRC-2, respectively, including the X-ray emission. In \S~\ref{DRC2_spectrum}, we investigate the MUSE datacube to extract the UV spectrum of DRC-2 and search for high-ionization emission lines, which are indicators of nuclear activity in this source.
In \S~\ref{DRC2_LAB} we discuss the possible link between the hidden QSO in DRC-2 and the nearby LAB (see Fig.~\ref{Fig_DRC2}). Finally, in \S~\ref{AGN_content} we compare the AGN content of the DRC with other protoclusters at lower redshifts.

\subsection{Spectral energy distribution and the X-ray-to-UV emission ratio of DRC-1}\label{DRC1_SED}
\begin{figure*}
	\begin{center}
		
		\includegraphics[width=90mm,keepaspectratio]{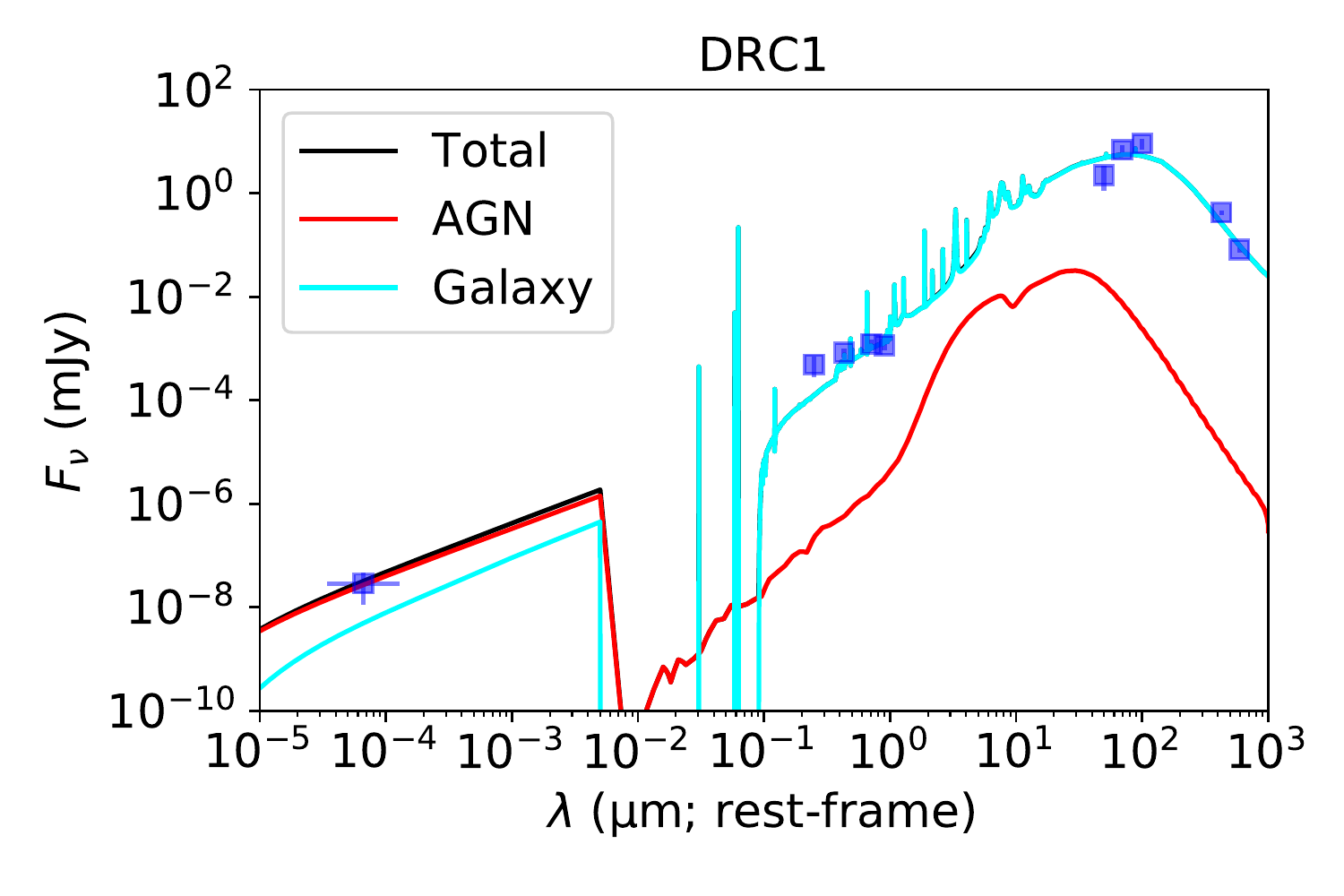}
		\includegraphics[width=90mm,keepaspectratio]{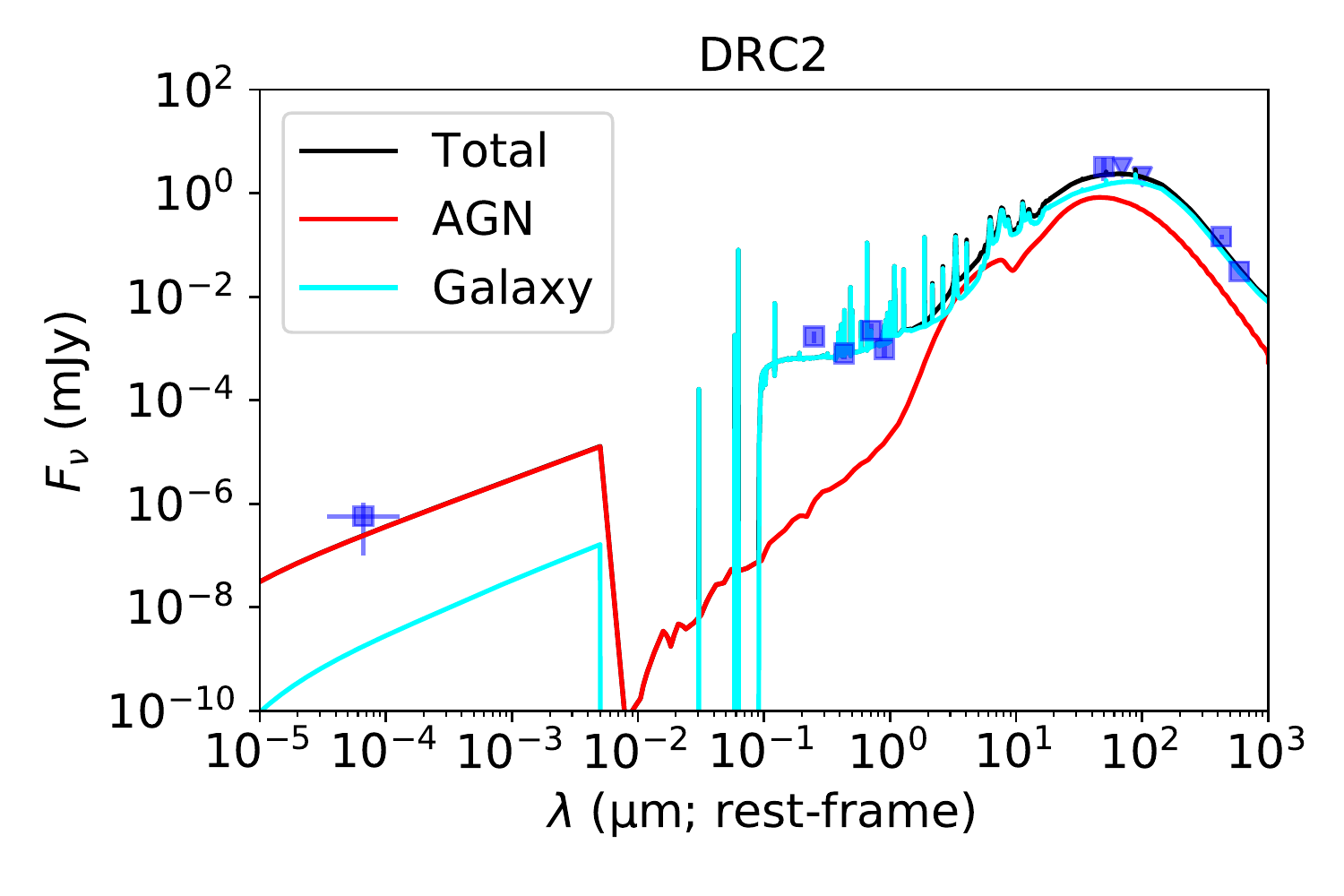}		
	\end{center}
	\caption{Best fitting SED of DRC-1 (\textit{left panel}) and DRC-2 (\textit{right panel}), decomposed into stellar (cyan line) and AGN (red line) components. Blue squares represent constrained photometric data, while downward-pointing triangles are upper limits.}\label{Fig_SED}
\end{figure*}

We performed SED fitting of DRC-1 using the X-CIGALE software \citep{Yang19}, which expands the CIGALE code \citep{Boquien19} to include a self-consistent treatment of X-ray photometry, allowing for the decomposition of the DRC-1 SED into AGN and galaxy components. We used similar parameter settings for the galaxy component as in \cite{Long20},
and added an AGN component as in \cite{Yang19} and \cite{Zou19}. 
We use the GEMINI/FLAMINGO-2 ($K_s$), HST WFC3 (F125W), \textit{Spitzer} IRAC ($3.5\,\mu m$ and $4.5\,\mu m$), \textit{Herschel}/SPIRE ($250\,\mu m$, $350\,\mu m$, $500\,\mu m$), and ALMA (2 mm and 3 mm) photometry reported in Tab. 1 of \cite{Long20}. We also added the absorption-corrected X-ray flux corresponding to the best-fitting absorbed power-law model described in \S~\ref{DRC-1}. The decomposed SED is shown in Fig.~\ref{Fig_SED}.

The resulting best-fitting stellar mass ($\approx2\times10^{11}\,\mathrm{M_\odot}$) and SFR ($\approx3400\mathrm{M_\odot\,yr^{-1}}$) are consistent with the results of \cite{Long20}. In addition, thanks to the X-ray data, we can investigate the AGN component, which is sub-dominant at all of the sampled wavelengths, except for the X-ray band. The AGN bolometric luminosity, log$\frac{L_{\mathrm{bol}}(\mathrm{AGN})}{\mathrm{erg\, s^{-1}}}\approx45.5$, is consistent with a bolometric correction $K_{\mathrm{bol}}=\frac{L_{\mathrm{bol}}}{L_X}$ of a few tens,
given the X-ray luminosity we derived in \S~\ref{DRC-1}, which is typical of moderately luminous QSOs \citep[e.g.][]{Lusso12,Duras20}. Having estimated of the AGN bolometric luminosity, we can derive the black-hole accretion rate as $\mathrm{BHAR}=\frac{(1-\epsilon)}{(\epsilon c^2)}L_{\mathrm{bol}}$, where we assumed a standard radiation efficiency $\epsilon=0.1$. We found $\mathrm{BHAR}\approx0.5\,\mathrm{M_\odot\,yr^{-1}}$.

X-CIGALE returns also the ratio between the X-ray and UV emission in QSOs, which  is usually parametrized with the quantity \hbox{$\alpha_{ox}=0.38\times \mathrm{log}(L_{2\,\mathrm{keV}}/L_{2500\angstrom})$} (e.g., \citealt{Brandt15} and references therein). $\alpha_{ox}$ carries information on the physical structure of the hot-corona and accretion-disk system in QSOs \citep[e.g.,][]{Lusso17}, and is well known to anti-correlate with the UV luminosity \citep[e.g.,][]{Steffen06,Just07,Lusso16}. The deviation of the measured value of $\alpha_{ox}$ from the expectation of the anti-correlation with UV luminosity is measured with the parameter \mbox{$\Delta\alpha_{ox}=\alpha_{ox}(\mathrm{observed}) - \alpha_{ox}(L_{2500\angstrom})$}. The best-fitting SED model of DRC-1 returns $\alpha_{OX}=-1.5$ and $\Delta\alpha_{OX}=0.03$; i.e., the X-ray-to-UV luminosity ratio of DRC-1 is well consistent with the predicted value.

\subsection{Spectral energy distribution and the X-ray-to-UV emission ratio of DRC-2}\label{DRC2_SED}

We repeated the SED fitting procedure with X-CIGALE for DRC-2, using the photometry reported in Tab. 1 of \cite{Long20}, and the absorption-corrected X-ray flux corresponding to the best-fitting MYTorus model described in \S~\ref{DRC-2}. The decomposed SED is shown in Fig.~\ref{Fig_SED}.
The resulting best-fitting stellar mass ($\approx4\times10^{10}\,\mathrm{M_\odot}$) and SFR ($\approx1250\,\mathrm{M_\odot\,yr^{-1}}$) are in agreement with the results of \cite{Long20}. 
X-CIGALE returns an AGN bolometric luminosity of log$\frac{L_{bol}(\mathrm{AGN})}{\mathrm{erg\, s^{-1}}}\approx46.5$.
In contrast with this value, scaling  the X-ray luminosity derived in \S~\ref{DRC-2} with the bolometric corrections of \cite{Lusso12} and \cite{Duras20}, we would have expected log$\frac{L_{bol}(\mathrm{AGN})}{\mathrm{erg\, s^{-1}}}\approx47.0$.
 We estimate the accretion rate of DRC-2 in the range $\mathrm{BHAR}=4.7-17.3\,\mathrm{M_\odot\,yr^{-1}}$, depending whether the bolometric luminosity derived from SED fitting or scaled from the X-ray emission is assumed.

The discrepancy between the two estimates of the bolometric luminosity can be explained if the QSO hosted in DRC-2 is particularly \mbox{X-ray} bright with respect to its intrinsic UV emission. 
The best-fitting SED model of DRC-2 returns $\alpha_{ox}=1.4$ and \mbox{$\Delta\alpha_{ox}=0.23$}.
This value of $\Delta\alpha_{ox}$ corresponds to a factor of $\approx4$ times higher X-ray luminosity than the expectation from the $L_X-L_{UV}$ relation, and is significantly higher than the value expected for typical radio-quiet QSOs (i.e., \mbox{$\Delta\alpha_{ox}=0$}), but still consistent with the upper bound of the \mbox{$\Delta\alpha_{ox}$} distribution of the radio-quiet QSO population \citep[e.g.,][]{Gibson08, Vito19b}. 

Enhanced X-ray emission is often observed in radio-loud QSOs \citep[e.g.,][]{Miller11, Ballo12}, due to the contribution of the jets to the total X-ray emission via inverse Compton scattering of external photons or synchrotron self-Compton scattering \citep[e.g.,][]{Medvedev20,Worrall20}. DRC-2 is covered with VLA (28 GHz) and ATCA (5.5 GHz) observations, but it is not detected \citep{Oteo18}. However, radio jets can have steep radio spectra, and may remain undetected at the high rest-frame frequencies probed by such data. Additional sensitive radio observations at lower frequencies are required to confirm the presence of radio jets in DRC-2.

Another possible explanation for the discrepancy between the bolometric luminosity derived from SED fitting and that inferred directly from the X-ray emission is related to the fitted \textit{Herschel} fluxes. These have been derived by \cite{Long20} with a thorough and complex deblending procedure using the ALMA 2 mm fluxes as priors, as the PSF of \textit{Herschel} is too large to resolve the individual DRC members. We investigated the effect of an underestimate of the far-IR fluxes produced by DRC-2. In particular, we repeated the SED fitting procedure under the extreme assumption that the entire DRC fluxes in the \textit{Herschel} bands were associated with DRC-2. We found log$\frac{L_{bol}(\mathrm{AGN})}{\mathrm{erg\,s^{-1}}}\approx47$, which is consistent with the luminosity inferred from X-ray data only. We also note that DRC-2 is resolved into multiple components by HST, but the used photometry refers to the entire system.

\subsection{UV spectrum of DRC-2}\label{DRC2_spectrum}

\begin{figure}
	\begin{center}
		
		\includegraphics[width=90mm,keepaspectratio]{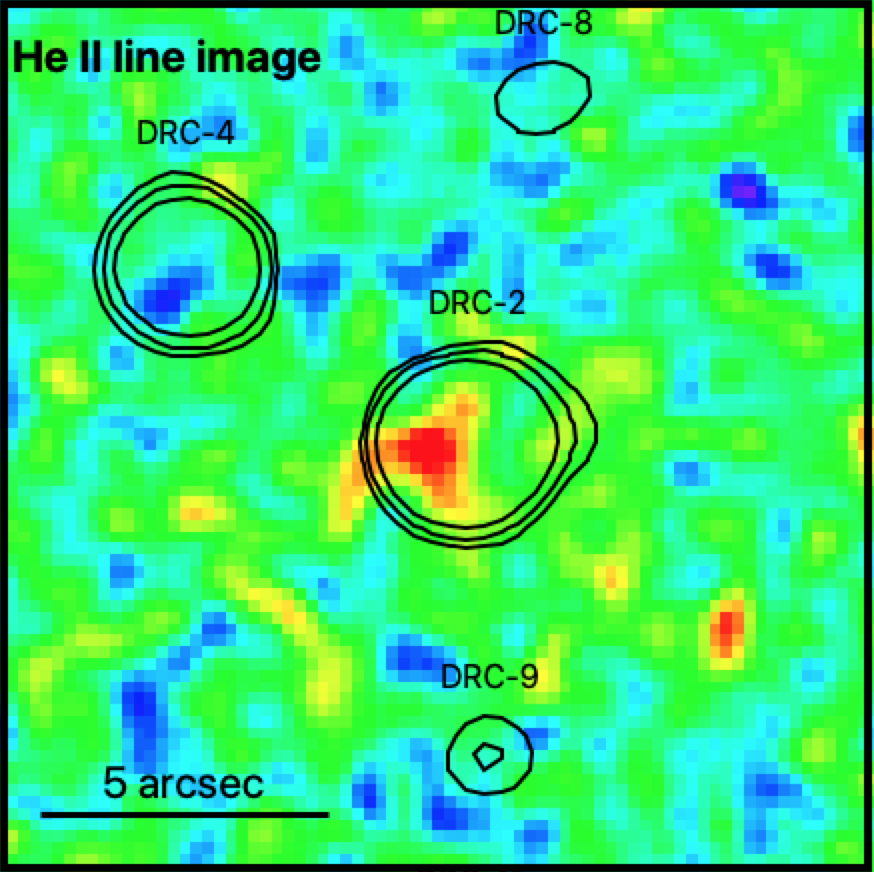}
	\end{center}
	\caption{Continuum-subtracted and smoothed MUSE datacube at $\mathrm{\lambda=8190-8210\,\mathrm{\AA}}$, where the He II emission line is expected to be found at $z=4.002$ (i.e., rest-frame $\lambda=1640\,\mathrm{\AA}$). Black contours mark the ALMA positions of the protocluster members.}   \label{Fig_DRC2_HeII}
\end{figure}

We investigated the MUSE datacube \citep{Oteo18} at the position of DRC-2 to search for high-ionization emission lines, like C IV $\lambda1549$ and He II $\lambda1640$. The presence of these features requires a strong and hard UV continuum, such as that produced by nuclear accretion. We found significant emission at $8200\,\mathrm{\AA}$ (Fig.~\ref{Fig_DRC2_HeII}) at $\approx0.5''$ from the ALMA and \chandra positions of DRC-2 (see Fig.~\ref{Fig_DRC2}), which we identify with the He II emission line redshifted at $z=4$. The He II line can also be produced by extremely young ($\lesssim10^7\,\mathrm{yr}$) Pop III stars \citep[][]{Scharer03} or metal-poor HMXBs \citep[e.g.][]{Eldridge12,Schaerer19,Stanway20}. However, as we noted also for DRC-1 in \S~\ref{DRC-1}, these are dusty (i.e., already enriched with metals) and massive galaxies, which are unlikely to be forming metal-poor stars. Therefore, the ionizing photons required for He~II line emission in this galaxy are most likely associated with the nuclear accretion we found in \S~\ref{DRC-2}. 

\begin{figure}
	\begin{center}
		
		\includegraphics[width=90mm,keepaspectratio]{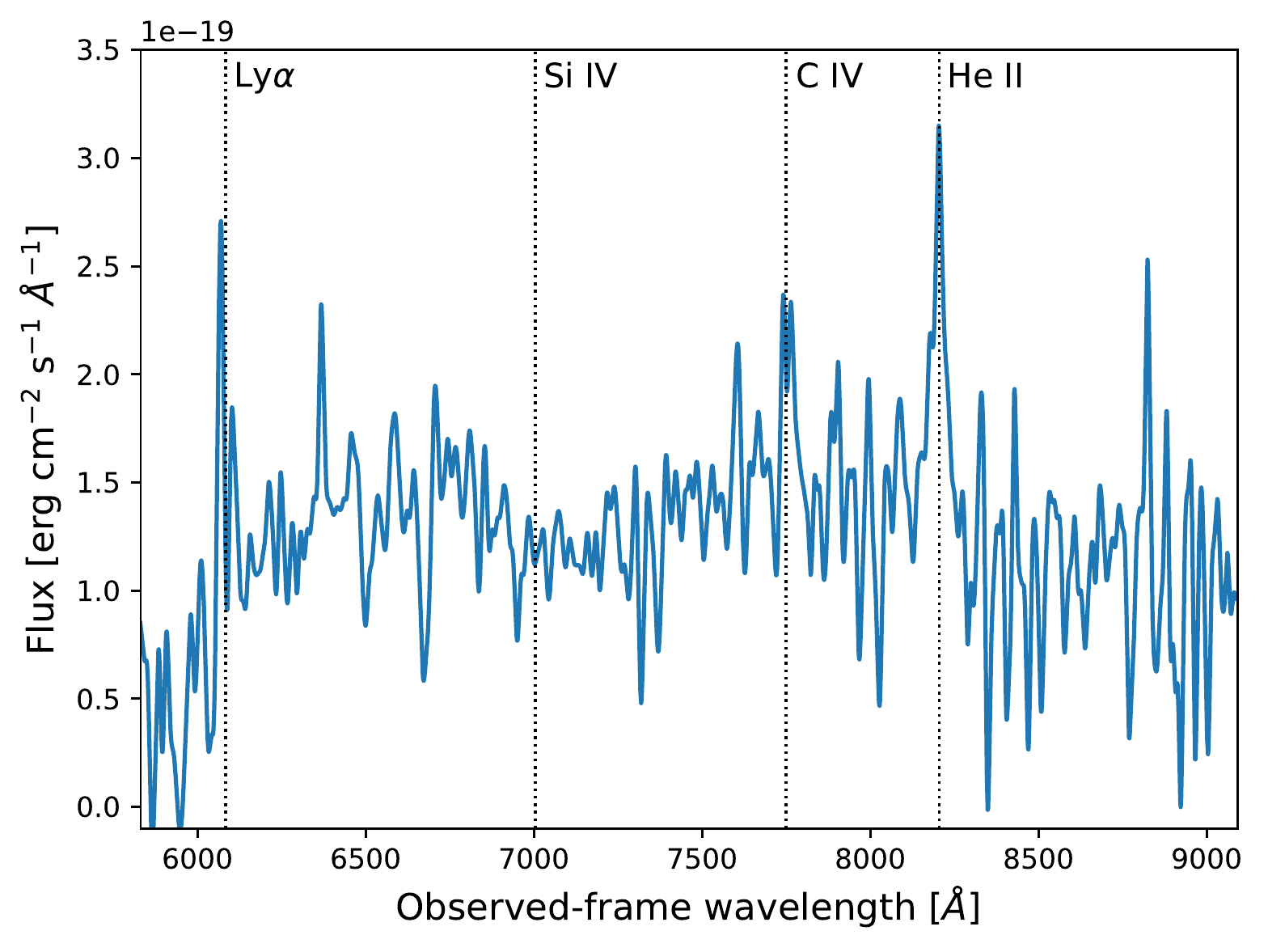}
	\end{center}
	\caption{Background-subtracted MUSE spectrum of DRC-2 extracted at the position of the He II emission peak. Vertical dashed lines mark the expected wavelengths of strong AGN features   redshifted at $z=4.002$.} \label{Fig_DRC2_spectrum}
\end{figure}

Fig.~\ref{Fig_DRC2_spectrum} presents the MUSE spectrum at the position of the He II emission line in DRC-2. We mark the expected wavelengths at $z=4.002$ of some of the strongest AGN emission features with vertical dashed lines. Besides the strong He II line, we also identify $Ly\alpha$ $\lambda1216$ emission and a drop in the faint continuum emission expected at rest-frame $\lambda<1216\,\mathrm{\AA}$ due to the $Ly\alpha$ forest and the Lyman break. It is unclear if the $Ly\alpha$ line in Fig.~\ref{Fig_DRC2_spectrum} is produced by the hidden QSO or if it is due to contamination by the LAB, since the extraction region is located at the border of the extended $Ly\alpha$ emission. We also note that the C IV line can  be identified in the MUSE spectra, although it falls in a spectral range with strong noise residuals. The detected UV emission could also be due to young and hot stars, although the presence of the C IV and He II lines point toward an AGN origin. We do not find strong evidence for extended C IV or He II emission in the LAB. 

Similar heavily obscured, extremely luminous QSOs with ``leaking" UV emission are the hot dust-obscured galaxies (Hot DOGs) presented by \cite{Assef15,Assef19} in which, according to the authors, the emission from the hidden central QSO is scattered by gas and dust into the line of sight. This scenario helps to explain the weakness of the UV emission relative to the intrinsic QSO power derived by SED decomposition and X-ray observations in those Hot DOGs. Similar physical processes may be in place in DRC-2, which has an estimated intrinsic luminosity similar to the Assef et al. QSOs.

\subsection{DRC-2 as the powering source of the LAB}\label{DRC2_LAB}
LABs  trace large reservoirs of cold gas, possibly flowing through cosmic filaments and fueling both star formation and SMBH accretion \citep[e.g.,][]{Umehata19}. 
\cite{Oteo18} reported the detection with MUSE of an LAB (with size $\approx60$ kpc) in the central part of the DRC, close to the position of DRC-2 (see Fig.~\ref{Fig_field} and \ref{Fig_DRC2}). The projected distance between the X-ray source centroid and the peak of the $Ly\alpha$ extended emission is $\lesssim1.9''$, corresponding to $\lesssim13.5$ kpc at $z=4$.
The heavily obscured and luminous QSO hiding in DRC-2, as revealed by \chandra, is the most obvious source to power the extended $Ly\alpha$ emission (e.g., \citealt{Geach09,Overzier16}), although independent mechanisms cannot be ruled out (e.g., ``cooling radiation"; e.g., \citealt{Haiman00}). The LAB luminosity (log$L_{Ly\alpha}\approx4\times10^{43}\,\mathrm{erg\,s^{-1}}$) is in line with other LABs powered by radio-quiet obscured QSOs \citep[e.g.,][]{Geach09,Overzier13,Borisova16}. 
Extended $Ly\alpha$ emission powered by QSOs can be produced by photoionization due to a strong ionizing flux of the central source or by resonant scattering of the $Ly\alpha$  photons emitted in the QSO emission line regions.

\begin{figure}
	\begin{center}
		
		\includegraphics[width=90mm,keepaspectratio]{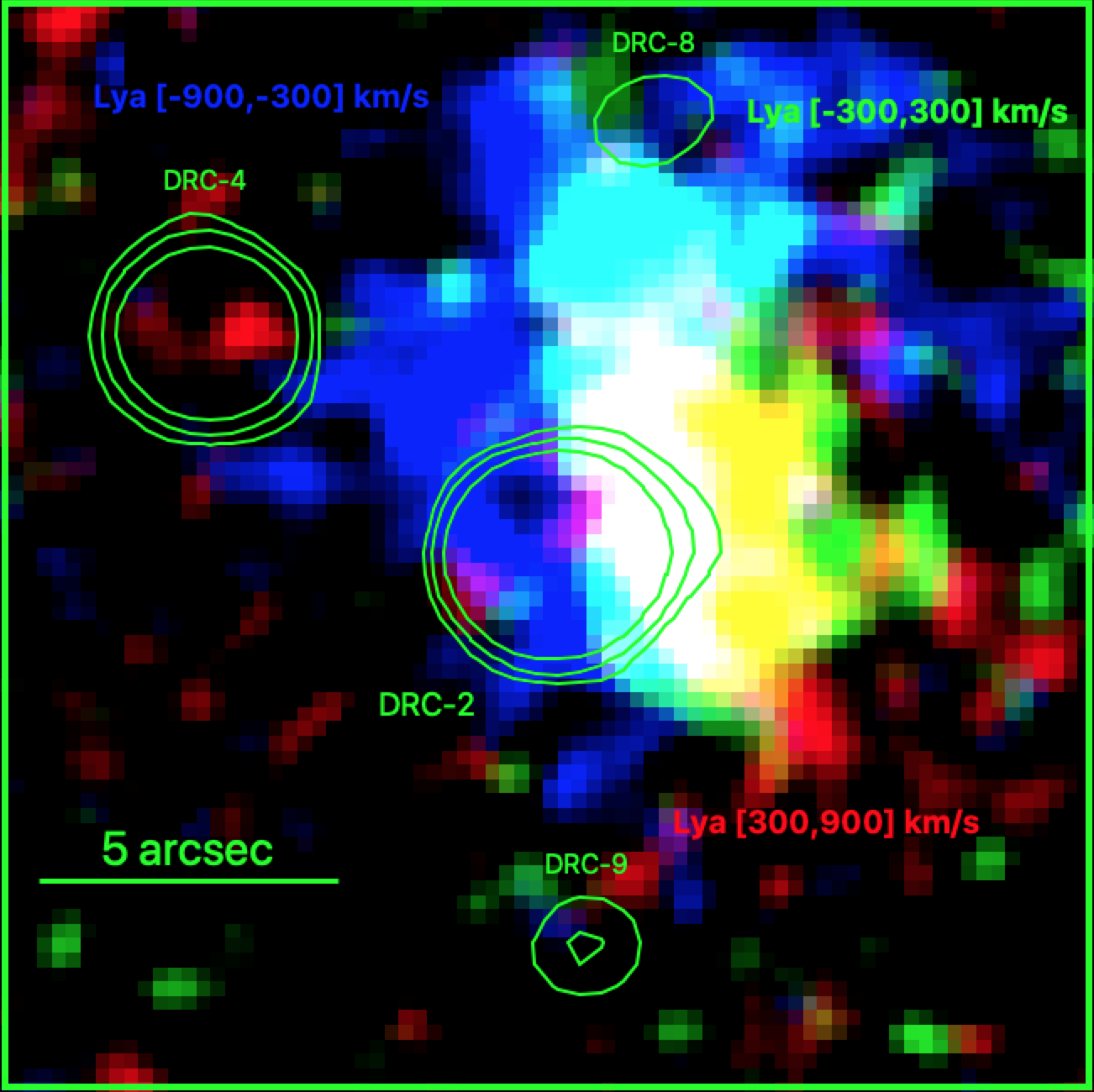}
		
	\end{center}
	\caption{Continuum-subtracted and smoothed $Ly\alpha$ emission in the velocity range $[-900,-300]\,\mathrm{km\,s^{-1}} $ (blue), $[-300,300]\,\mathrm{km\,s^{-1}} $ (green), and $[+300,+900]\,\mathrm{km\,s^{-1}} $ (red). Zero velocity is assumed to be at $z=4.002$. Green contours mark the ALMA positions of the protocluster members.}\label{Fig_LAB_vel}
\end{figure}

Another intriguing, but currently quite speculative,  possibility is that the extended $Ly\alpha$ emission traces gas shocked by a QSO-driven or starburst-driven outflow. \cite{Oteo18} showed that the profile of the CO(4--3) emission line in DRC-2 is strongly asymmetric and double peaked ($\Delta v\approx1000\,\mathrm{km\,s^{-1}}$). Moreover, the ALMA continuum emission (with an angular resolution of $\approx2"$) is elongated toward the projected position of the maximum of the extended $Ly\alpha$ intensity (see green and cyan contours in Fig.~\ref{Fig_field} and \ref{Fig_DRC2}). While these findings could be related to the multiple components revealed with HST contributing to the total ALMA flux with different velocities, they may also be due to outflowing material. The extended $Ly\alpha$ emission also appears to be structured in velocity space (Fig.~\ref{Fig_LAB_vel}). In particular, the $Ly\alpha$ emission at the position of maximum intensity is redshifted with respect to the outer parts of the LAB. The double peaked CO(4--3) emission line and the positional coincidence of the tip of the elongated ALMA emission with the maximum $Ly\alpha$ surface brightness may be indicative of the presence of a massive dusty and molecular outflow ejected from the DRC-2 nuclear regions. The mechanical push of the ouflowing material on the gas can also explain the redshifted $Ly\alpha$ emission at the impact point. However, most of these properties can also be ascribed to projection effects, and an inflow scenario cannot be excluded. New ALMA spectroscopic observations of molecular lines with higher sensitivity and angular resolution ($\lesssim1"$) will securely identify a possible massive outflow, allowing us to assess its geometry and its possible contribution to the powering of the LAB. Moreover, deeper MUSE observations of the LAB will study in detail its geometry and velocity structure.

\subsection{AGN content in the structure}\label{AGN_content}

In order to compare the AGN content of the DRC protocluster with similar structures at lower redshifts, we considered only AGN identified in sub-mm galaxies (SMGs), which were confirmed spectroscopically to belong to each structure.
Among the 13 spectroscopically identified members of the DRC protocluster, we detect two X-ray AGN (DRC-1 and DRC-2). \cite{Oteo18} reported that DRC-6, which is not detected in our \textit{Chandra} observations, is a radio galaxy (see \S~\ref{intro}). Therefore DRC-6 likely hosts a low-luminosity and/or obscured AGN.
Thus, the fraction of known AGN among DSFGs in this protocluster is $23^{+22}_{-12}\%$. Considering only X-ray detected sources, the fraction is $15^{+20}_{-10}\%$, which is consistent with the X-ray selected AGN fraction in field SMGs \citep[e.g.,][]{Wang13_1}. 
However, as discussed in \S~\ref{stack}, some of the DRC protocluster members undetected in our \textit{Chandra} observations may host low-luminosity AGN, resulting in a higher intrinsic AGN fraction.

We compare these values with the results found for SSA22, a massive protocluster at $z=3.09$. Considering only the 8 SMGs detected at 1.1 mm in this structure \citep{Umehata15},  4 of them ($50^{+39}_{-24}\%$) are X-ray detected AGN \citep{Lehmer09a}.\footnote{A ninth SMG in SSA22 which lacks spectroscopic confirmation but is associated with a photometric redshift of $z=3.08$ (i.e., a likely member of SSA22) is also detected in the X-rays. } Although the 400 ks \textit{Chandra} observation of SSA22 at $z\approx3.1$ reaches a deeper flux limit and intrinsic luminosity limit than our observation of the DRC, all of the X-ray AGN among SMGs in SSA22 have fluxes $F_{0.5-7\,\mathrm{keV}}>10^{-15}\,\mathrm{erg\,cm^{-2}\,s^{-1}}$, and could have been detected in our \textit{Chandra} observation. Therefore, in this sense, the comparison is fair. A caveat instead arises from the selection of SMGs in the two structures, as 1 mm observations have been used for SSA22 at $z=3.1$, while SMGs in the DRC structure at $z=4$ have been selected in 2 mm and 3 mm observations.

Another well-studied, massive protocluster is the structure at $z=2.3$ discovered by \cite{Steidel05} in the field of the QSO HS 1700+643.
Using 200 ks \textit{Chandra} observations, \cite{Digby-North10} did not find a significant enhancement of X-ray selected AGN among galaxy members of that protocluster. Recently, \cite{Lacaille19} presented SCUBA-2 observations of the same field at $850\mu$m and $450\mu$m.
In order to estimate the AGN content among the 4 spectroscopically confirmed sub-mm galaxies in the HS 1700+643, we matched their coordinates with the positions of X-ray sources in the \textit{Chandra} Source Catalog 2.0,\footnote{\url{https://cxc.harvard.edu/csc/}} finding no match. This result corresponds to an AGN fraction among sub-mm galaxies of $<27\%$.

From the comparison discussed in this section, we conclude that the AGN fraction in sub-mm galaxies in the DRC protocluster is not significantly different from those in similar structures at lower redshift, although the statistics upon which this result is based are limited. Future observations of the DRC protocluster aimed at identifying $Ly\alpha$ emitter and Lyman-break galaxies will allow the investigation of the AGN content in the more general galaxy population, and comparison of the results with a larger sample of similar structures for which sub-mm/mm observations are currently lacking (e.g., \citealt{Lehmer09b,Digby-North10,Charutha17}, but see also \citealt{Macuga19}).

Considering the BHAR values described in \S~\ref{DRC1_SED} and \ref{DRC2_SED}, BH accretion in the DRC is strongly dominated by DRC-2, which accretes $\gtrsim10$ times faster than DRC-1, the second most luminous AGN in the structure. As discussed above, DRC-6 likely hosts a low-luminosity AGN, thus most probably characterized by a low BHAR. From the stacking analysis performed in \S~\ref{stack}, the typical BHAR of individually undetected DRC members is $\approx0.1\,\mathrm{M_\odot\,yr^{-1}}$, and thus their cumulative accretion rate is comparable to, or even higher than, the BHAR of DRC-1.

\section{Conclusions}

We presented new \chandra observations (139 ks) of the DRC, a starbursting and gas-rich protocluster, constituted of at least 13 DSFGs at $z=4$. We summarize here the main results.
\begin{itemize}
	\item We detected in the X-ray band two members of the protocluster, DRC-1 and DRC-2. Both sources present a hard \mbox{X-ray} spectrum, consistent with being heavily obscured. These two SMGs have the highest gas content among the protocluster members \citep{Oteo18}. Moreover, high-resolution ALMA and HST imaging resolved them into multiple likely interacting star-forming clumps. This is consistent with the idea that the availability of a large amount of gas and galaxy interactions, both of which are enhanced in gas-rich, unvirialized overdensities at high redshift, can trigger fast and obscured SMBH accretion.
	
	\item DRC-1 is detected in the full and hard X-ray bands. From X-ray hardness ratio arguments and basic spectral analysis we estimated a high obscuration level ($N_H\approx10^{24}\,\mathrm{cm^{-2}}$) and moderate intrinsic luminosity
	(\mbox{$L_{2-10\,\mathrm{keV}}\approx2\times10^{44}\,\mathrm{erg\,s^{-1}}$}).
	
	\item DRC-2 is a bright hard X-ray source \mbox{($F_{2-7\mathrm{keV}}\approx3\times10^{-15}\,\mathrm{erg\,cm^{-2}\,s^{-1}}$)}, while it is not detected in the soft band. We fitted its X-ray spectrum with the MYTorus model and found a best-fitting column density $N_H>10^{24}\,\mathrm{cm^{-2}}$ and intrinsic luminosity \mbox{$L_{2-10\,\mathrm{keV}}\approx3\times10^{45}\,\mathrm{erg\,s^{-1}}$}. According to these values, DRC-2 hosts a powerful, Compton-thick QSO candidate, with an X-ray luminosity similar to the most-luminous known optically selected type 1 QSOs. 
	
	\item We fitted the SEDs of DRC-1 and DRC-2 and decomposed them using galaxy and AGN templates. Resulting stellar masses and star-formation rates are consistent with previous findings. Adding the X-ray data, we could investigate for the first time the AGN component in these two galaxies. The bolometric luminosity of the QSO hosted in DRC-1 is in line with the expectation from the X-ray emission. In contrast, the bolometric luminosity of the QSO in DRC-2 derived via SED fitting is a factor of $\approx10$ lower than the estimate from X-ray data. 
	While this discrepancy has been found in other radio-quiet QSOs, it could be explained by the presence of radio jets in DRC-2, undetected by current radio observations due to the high rest-frame frequencies that they sample.
	
	\item Led by the discovery of a luminous obscured QSO in DRC-2, we investigated previously performed  MUSE observations to search for high-ionization emission lines consistent with the position of DRC-2. We derived the rest-frame UV spectrum, detecting the continuum, strong He II and $Ly\alpha$ emission lines, and possibly the C IV emission line. Due to the presence of high-ionization emission lines, we consider the UV emission as of nuclear origin. We speculate that the detection of UV continuum and emission lines in a DSFG hosting a Compton-thick QSO might be due to UV photons scattered back into the line of sight, as found in other luminous and obscured QSOs.

	\item A $Ly\alpha$ blob (LAB) was detected by \cite{Oteo18} with MUSE observations in the proximity of DRC-2. The luminous obscured QSO hosted in this SMG is the most obvious source for powering the LAB, through photoionization or scattering of $Ly\alpha$ photons. This implies a largely unobscured line of sight in the direction of the LAB. Alternatively, we propose that DRC-2 could be launching a powerful dusty and molecular outflow that ionized the surrounding gas through shocks, producing the LAB. This interpretation is supported by the highly asymmetric and doubled peaked profile ($\Delta v_{peak}\approx1000\,\mathrm{km\,s^{-1}}$) of the CO(4-3) emission line \citep{Oteo18}. Moreover, ALMA revealed that the DRC-2 dust emission is elongated toward the LAB, and reaches the peak position of the extended $Ly\alpha$ surface brightness.

	\item 	Considering also a known radio galaxy in another member of the structure, DRC-6, in addition to DRC-1 and DRC-2, the AGN content among SMGs in this protocluster ($23^{+22}_{-12}\%$) is consistent with the fractions estimated for other protoclusters at lower redshifts ($z\approx2-3$). However, an X-ray stacking analysis reveals that nuclear accretion producing X-ray emission below our detection threshold might be ongoing also in other SMG members of the DRC.
\end{itemize}

\begin{acknowledgements}
	We thank the anonymous referee for their prompt and useful comments.
	We thank E. Vanzella for his help with MUSE data, and A. Long for providing ALMA and HST contours.
FV acknowledges financial support from CONICYT and CASSACA through the Fourth call for tenders of the CAS-CONICYT Fund. We acknowledge financial contribution from ANID grants Basal-CATA AFB-170002 (FV, FEB), FONDECYT Regular 1190818 (FEB) and1200495 (FEB), and the Ministry of Economy, Development, and Tourism's Millennium Science Initiative through grant IC120009, awarded to The Millennium Institute of Astrophysics, MAS (FEB). WNB and FZ acknowledge CXC grant GO9-20099X. RG acknowledges financial contribution from the agreement ASI-INAF n. 2017-14-H.O. CS is grateful for support from the National Research Council of Science and Technology, Korea (EU-16-001). This research has made use of data obtained from the Chandra Data Archive and software provided by the Chandra X-ray Center (CXC) in the application package CIAO. Based on observations collected at the European Organisation for Astronomical Research in the Southern Hemisphere under ESO programme 295.A-5029(A).
Images have been produced with SAO Image DS9 \citep{Joye03}. This research made use of Astropy,\footnote{http://www.astropy.org} a community-developed core Python package for Astronomy \citep{Astropy13, Astropy18}

\end{acknowledgements}

%
%
\bibliographystyle{aa}
\bibliography{biblio.bib} 

%

\appendix
\section{X-ray images of undetected protocluster members}\label{Appendix_undetect}
\begin{figure*}
	\begin{center}
		
		\includegraphics[width=160mm,keepaspectratio]{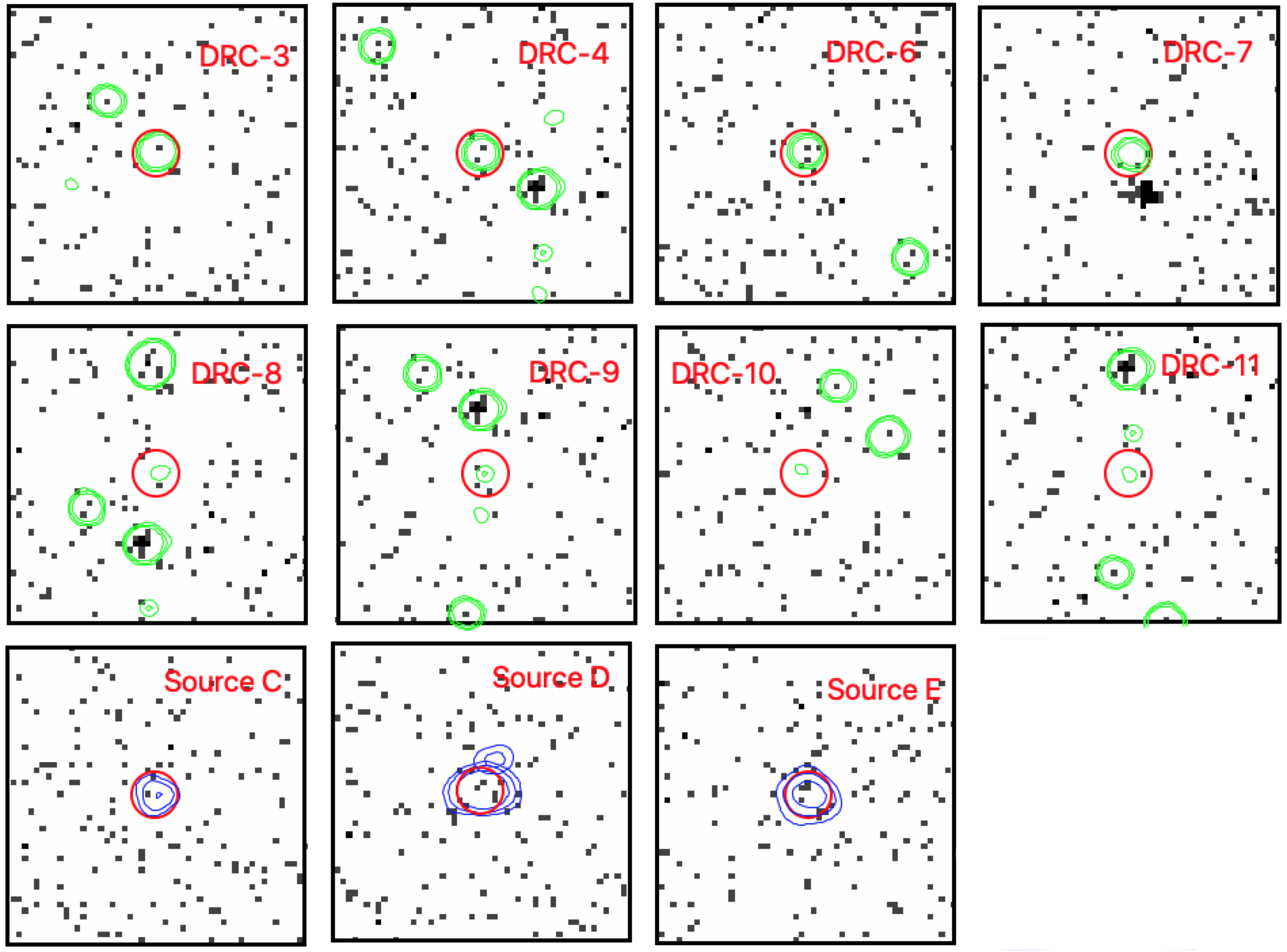}
		
	\end{center}
	\caption{Full-band $25''\times25''$ images of the individually undetected, spectroscopically confirmed members of the protocluster. The $R=2''$ red circles are centered at the ALMA positions. Green contours mark the continuum emission at 2 mm of the sources in the core of the structure, as in Fig.~\ref{Fig_field}. Sources in the outskirts of the protoclusters are marked with CO(4-3) contours ($3\sigma$, $5\sigma$, and $9\sigma$), since most of them are not associated with continuum emission \citep{Ivison20}. In the field of source D, a second line emitter ($\approx3''$ from the primary source; $\Delta v \approx1700\,\mathrm{km\,s^{-1}}$) was found and discussed by \citet{Ivison20}. It is not clear if this is another protocluster member, or if it is part of source D (e.g., an outflowing component).}\label{Fig_undetect}
\end{figure*}

Fig.~\ref{Fig_undetect} presents the full-band images of the spectroscopically confirmed members of the protocluster which are not detected individually (see \S~\ref{results}).
 
\section{X-ray emission in unidentified field ALMA sources}\label{Appendix_unident}
\begin{figure*}
	\begin{center}
		
		\includegraphics[width=160mm,keepaspectratio]{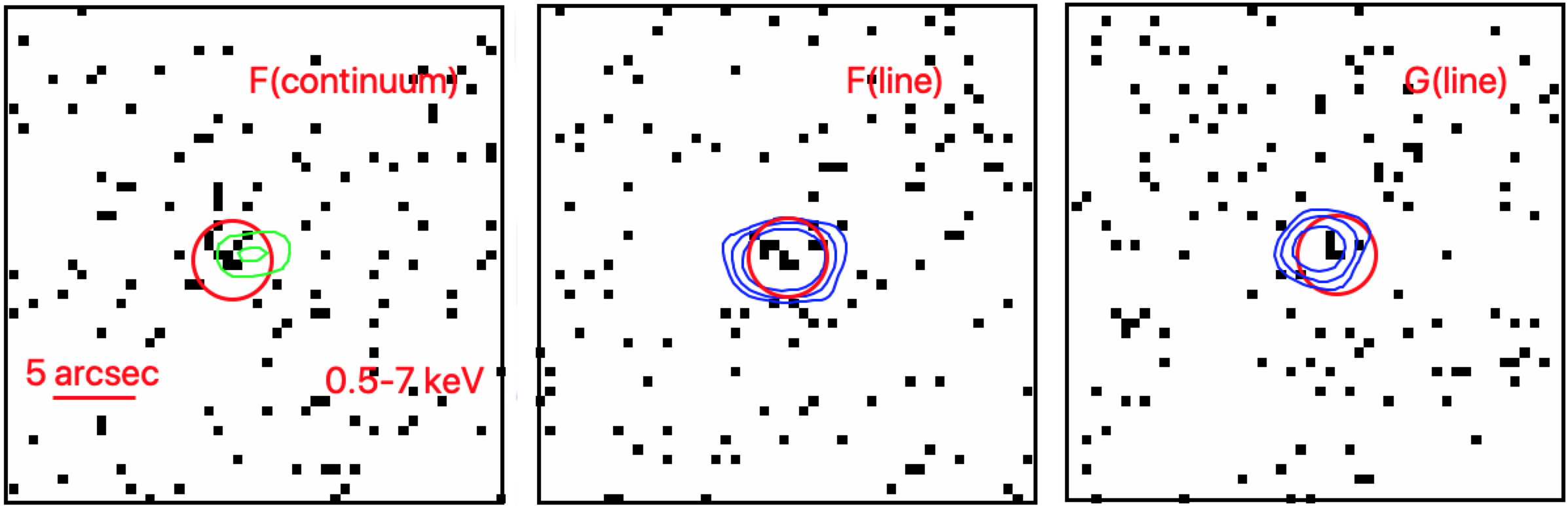}
		
	\end{center}
	\caption{Full-band $25''\times25''$ images of the X-ray detected, spectroscopically unidentified ALMA sources in the outskirts of the DRC field \citep{Ivison20}. The $R=2"$ red circles are centered on the X-ray positions. The green contours mark the 3 mm emission of the continuum source in the field of sub-mm source F. Blue contours mark the CO(4-3) line emission for the unidentified line emitters (not matched with continuum emission) in the fields of sub-mm sources F and G (see \citealt{Ivison20} for details).}\label{Fig_unident}
\end{figure*}

Besides the 13 spectroscopically confirmed members of the protocluster, several additional LABOCA 870$\mu$m sources have been detected by ALMA in the outskirts of the DRC field (see \citealt{Ivison20}, for details). Most of them have been resolved with ALMA into multiple galaxies, either (unidentified) line emitters or continuum emitters, or both.
In this section, we briefly discuss three of these sources which are associated with detected X-ray emission. In particular, we detected significant X-ray emission ($1-P_B>0.99$) associated with a 3 mm continuum source and a line emitter in the field of LABOCA source F, which we refer to as sources F$^{cont}$ and F$^{line}$, respectively, and a line emitter in the field of LABOCA source G, which we refer to as G$^{line}$ (Fig.~\ref{Fig_unident}). 
X-ray photometry has been computed in a circular region centered on the centroid of the X-ray sources.

Source F$^{cont}$ is detected in the hard band with $5.6_{-2.1}^{+2.8}$ net counts ($F_{2-7\,\mathrm{keV}}=4.0_{-1.5}^{+2.0}\times10^{-16}\,\mathrm{erg\,cm^{-2}\,s^{-1}}$), while no count has been detected in the soft band ($<1.1$ net counts; $F_{0.5-2\,\mathrm{keV}}<8\times10^{-17}\,\mathrm{erg\,cm^{-2}\,s^{-1}}$), corresponding to a hardness ratio of $HR>0.40$ and an effective photon index of $\Gamma_{eff}<0.29$. Source F$^{line}$ is detected both in the soft and hard bands, with $3.7_{-2.4}^{+1.7}$ ($F_{0.5-2\,\mathrm{keV}}=2.7_{-1.7}^{+1.2}\times10^{-16}\,\mathrm{erg\,cm^{-2}\,s^{-1}}$) and  $6.6_{-3.0}^{+2.3}$ ($F_{2-7\,\mathrm{keV}}=4.7_{-2.2}^{+1.7}\times10^{-16}\,\mathrm{erg\,cm^{-2}\,s^{-1}}$) net counts, respectively,  corresponding to $HR=0.23^{+0.26}_{-0.31}$ and  $\Gamma_{eff}=0.69^{+0.67}_{-0.64}$. Source G is detected only in the hard band with $3.5^{+2.4}_{-1.7}$ net counts ($F_{2-7\,\mathrm{keV}}=2.5_{-1.2}^{+1.7}\times10^{-16}\,\mathrm{erg\,cm^{-2}\,s^{-1}}$), while $<2.1$ net counts are detected in the soft band ($F_{0.5-2\,\mathrm{keV}}<1.5\times10^{-16}\,\mathrm{erg\,cm^{-2}\,s^{-1}}$), corresponding to $HR>0.28$ and $\Gamma_{eff}<0.57$. Therefore, all of these sources present hard X-ray spectra, probably due to ongoing nuclear accretion in obscured conditions.

\end{document}